\documentclass[12pt]{article}
\usepackage{epsf}

\usepackage{epsfig,graphics}
\usepackage {graphicx}
\usepackage {epsfig}
\usepackage {subfigure}
\usepackage {tabularx} 
\usepackage{rotate}	
\usepackage{slashed}

\usepackage{amsmath}
\usepackage{amsfonts}
\usepackage{amssymb}
\usepackage{graphicx}
\usepackage{cite}

\newcommand{\bmat}{\left(\begin{array}}
\newcommand{\emat}{\end{array}\right)}

\def\yzero{\smash{\hbox{$y\kern-4pt\raise1pt\hbox{${}^\circ$}$}}}

\def\beq{\begin{equation}}
\def\eeq{\end{equation}}
\def\beqa{\begin{eqnarray}}
\def\eeqa{\end{eqnarray}}

\def\-{\hphantom{-}}

\def\s2{\frac{1}{\sqrt2}}

\def\beq{\begin{equation}}
\def\eeq{\end{equation}}
\def\beqa{\begin{eqnarray}}
\def\eeqa{\end{eqnarray}}

\def\IF{\relax{\rm I\kern-.18em F}}
\def\II{\relax{\rm I\kern-.18em I}}
\def\IP{\relax{\rm I\kern-.18em P}}
\def\IC{\relax\hbox{\kern.25em$\inbar\kern-.3em{\rm C}$}}
\def\IR{\relax{\rm I\kern-.18em R}}

\def\Dsl{\,\raise.15ex\hbox{/}\mkern-13.5mu D} %this one can be subscripted
\def\IZ{Z\kern-.4em  Z}

\catcode`\@=11   
\newdimen\@rotdimen
\newbox\@rotbox  

\def\@vspec#1{\special{ps:#1}}%  passes #1 verbatim to the output
\def\@rotstart#1{\@vspec{gsave currentpoint currentpoint translate
   #1 neg exch neg exch translate}}% #1 can be any origin-fixing transformation
\def\@rotfinish{\@vspec{currentpoint grestore moveto}}% gets back in synch 
\def\@rotr#1{\@rotdimen=\ht#1\advance\@rotdimen by\dp#1%
   \hbox to\@rotdimen{\hskip\ht#1\vbox to\wd#1{\@rotstart{90 rotate}%
   \box#1\vss}\hss}\@rotfinish}
\def\@rotl#1{\@rotdimen=\ht#1\advance\@rotdimen by\dp#1%
   \hbox to\@rotdimen{\vbox to\wd#1{\vskip\wd#1\@rotstart{270 rotate}%
   \box#1\vss}\hss}\@rotfinish}%
\def\@rotu#1{\@rotdimen=\ht#1\advance\@rotdimen by\dp#1%
   \hbox to\wd#1{\hskip\wd#1\vbox to\@rotdimen{\vskip\@rotdimen
   \@rotstart{-1 dup scale}\box#1\vss}\hss}\@rotfinish}%
\def\@rotf#1{\hbox to\wd#1{\hskip\wd#1\@rotstart{-1 1 scale}%
   \box#1\hss}\@rotfinish}%
\def\rotate{\@ifnextchar[{\@rotate}{\@rotate[l]}}
\def\@rotate[#1]#2{\setbox\@rotbox=\hbox{#2}\@nameuse{@rot#1}\@rotbox}

\catcode`\@=12
\topmargin
-1.5cm
\textwidth
15.5cm
\textheight
23.5cm
\oddsidemargin
0.7cm
\evensidemargin
0.7cm

\begin{document}

%----------------------------------------------------------------------%
%  numbering equations with section number
%----------------------------------------------------------------------%
\makeatletter
\@addtoreset{equation}{section}
\makeatother
\renewcommand{\theequation}{\thesection.\arabic{equation}}
%----------------------------------------------------------------------%
%  title page
%----------------------------------------------------------------------%
\pagestyle{empty}
%\vspace*{1.0in}
\vspace{-0.2cm}
\rightline{ IFT-UAM/CSIC-13-081}
\rightline{FTUAM-13-19}
%\rightline{\tt hep-th/xxxxxxx}
\vspace{1.2cm}
\begin{center}

%\vspace{0.5cm}

\LARGE{The String Origin of SUSY Flavor Violation} 
\\[13mm]
  \large{Pablo G. C\'amara$^{a}$,   Luis E. Ib\'a\~nez$^{b} $ and Irene  Valenzuela$^{b}$  \\[6mm]}
\small{
$^a$ Departament d'Estructura i Constituents de la Mat\`eria and Institut de Ci\`encies\\[-0.3em] 
del Cosmos, Universitat de Barcelona, Mart\'{\i} i Franqu\'es 1, 08028 Barcelona, Spain\\
 $^b$  Departamento de F\'{\i}sica Te\'orica
and Instituto de F\'{\i}sica Te\'orica UAM/CSIC,\\[-0.3em]
Universidad Aut\'onoma de Madrid,
Cantoblanco, 28049 Madrid, Spain 
\\[8mm]}
\small{\bf Abstract} \\[7mm]
\end{center}
\begin{center}
\begin{minipage}[h]{15.22cm}
We argue that in large classes of string compactifications with a MSSM-like structure substantial flavor violating 
SUSY-breaking soft terms are generically induced. 
We specify to the case 
of flavor dependent soft-terms  in  
type IIB/F-theory  SU(5) unified models, although our results can be easily extended to other settings. 
The Standard Model (SM) degrees of freedom reside in a local system of 7-branes wrapping a 4-fold $S$ 
in the extra dimensions. 
It is known that in the presence of closed 
string 3-form fluxes  
SUSY-breaking terms are typically generated. 
We explore the generation dependence of these soft
terms and find that non-universalities  
arise whenever the flux 
varies 
over the 4-fold $S$. 
These non-universalities are parametrically suppressed by 
$(M_{\rm GUT}/M_{\rm Pl})^{1/3}$.
They also arise in the case of varying open string fluxes,  
in this  case  parametrically suppressed by $\alpha_{\rm GUT}^{1/2}$.
For a standard unification scheme with $M_{\rm GUT}\simeq 10^{16}$ GeV and $\alpha_{\rm GUT}\simeq 1/24$
these suppressions are very mild. Although limits from the kaon mass difference $\Delta m_K$ 
are easily obeyed for squark masses 
above the present LHC limits, constraints from the CP-violation parameter $\epsilon_K$  imply 
squark masses in the multi-TeV region.  The constraints from BR($\mu\rightarrow e\gamma$)
turn out to be the strongest ones,  with 
slepton masses of order $\sim 10$ TeV or heavier required to obey the experimental limits.
These sfermion 
masses are
consistent with the observed large value $m_H\simeq 126$ GeV of the Higgs mass.
We discuss under what conditions such strong limits may be relaxed allowing for 
SUSY particle production at LHC.

\end{minipage}
\end{center}
\newpage
%----------------------------------------------------------------------%
%  Resetting of counters
%----------------------------------------------------------------------%
\setcounter{page}{1}
\pagestyle{plain}
\renewcommand{\thefootnote}{\arabic{footnote}}
\setcounter{footnote}{0}
%----------------------------------------------------------------------%
%  Paper begins
%----------------------------------------------------------------------%

%\end{document}

%\end{document}

\tableofcontents

%&&&&&&&&&&&&&&&&&&&&&&&&&&&&&&&&&
\section{Introduction}
%&&&&&&&&&&&&&&&&&&&&&&&&&&&&&&&&&

One of the most relevant aspects of low-energy supersymmetry is the 
potentially large contribution of supersymmetric particles to processes that involve Flavor Changing Neutral Currents (FCNC's). 
These include the  $K^0-{\overline K}^0$  oscillation and CP-violating parameters $\Delta m_K$ and $\epsilon _K$, as well as
lepton number violating transitions like $\mu\rightarrow e\gamma$ and other hadronic and leptonic processes involving 
heavier generations. All these transitions may be induced by SUSY particles in the presence of flavor changing 
SUSY-breaking soft parameters like off-diagonal scalar masses ${\tilde m}_{ij}$, $i\not=j$  
\cite{Masiero:1997bv,Paradisi:2005fk,Ciuchini:2007ha,Masina:2002mv,Chung:2003fi}
. These flavor violating 
contributions may  be too large and violate experimental bounds unless the squark masses of the first two generations are almost degenerated and 
off-diagonal masses are much suppressed. Otherwise the SUSY spectrum should be very heavy, effectively decoupling 
from these low-energy transitions.

The presence or not of flavor violating couplings depends, of course,  on what  the underlying source of SUSY-breaking is and on how  it is transmitted to the visible sector. In the context of gravity mediation,  off-diagonal 
flavor violating scalar masses may appear e.g. if the K\"ahler metric of the MSSM quark/lepton superfields and their derivatives 
on SUSY-breaking scalars do not align  in flavor space, see e.g.~\cite{Brignole:1997dp} and references therein. Similar flavor violating transitions may also originate from non-perturbative superpotential contributions \cite{Berg:2012aq}. 
Other schemes like gauge or anomaly mediation which transmit SUSY-breaking in a flavor universal manner were  put forward 
in order to avoid the presence of too large flavor violating transitions.

The question that we want to address in this work is whether in the  more fundamental setting of String Theory one can
obtain information about the presence of flavor violating SUSY-breaking soft terms. This is a question
which has been controversial and somewhat  author-dependent in the last two decades
(for some discussions on this issue see e.g. \cite{Brignole:1993dj,Louis:1994ht,Conlon:2007dw,Choi:2008hn,Camara:2011nj}.) 
The reason being that it is a general question which is difficult to answer 
without {\it both}  a scheme giving rise to a MSSM compactification as well as a tractable source of SUSY-breaking.

During the last fifteen  years important progress has been achieved both in the construction of MSSM-like compactifications 
as well as controlled  sources of SUSY-breaking in String Theory \cite{Ibanez:2012zz}. In particular it has been found that generic closed string 
fluxes in type IIB orientifold compactifications (or their F-theory generalisations) provide a tractable source of 
SUSY-breaking soft terms. Such fluxes can be topologically non-trivial or may arise instead from the backreaction of localized SUSY-breaking sources present in the compactification. In the simplest situations some Imaginary Self-Dual (ISD)  3-form flux $G_3$ may lead to non-trivial soft terms in a quite general setting. From the point of view of the effective 4d supergravity, SUSY-breaking is triggered 
by the auxiliary fields of the K\"ahler moduli ({\it moduli dominance}).  
Explicit expressions for the resulting soft terms 
have been obtained \cite{Grana:2002nq,Camara:2003ku, Grana:2003ek, Lust:2004fi, Camara:2004jj} and their low-energy LHC phenomenology has been analysed \cite{Aparicio:2008wh, Aparicio:2012iw, Aparicio:2012vk, Aparicio:2012ju} (see also   \cite{Choi:2005ge, Conlon:2006wz, Conlon:2007xv, Blumenhagen:2009gk, Heckman:2009bi, Acharya:2008zi, Kane:2011kj, Li:2011ab}). 

\begin{figure}[t]
\begin{center}
\includegraphics[width=8.0cm]{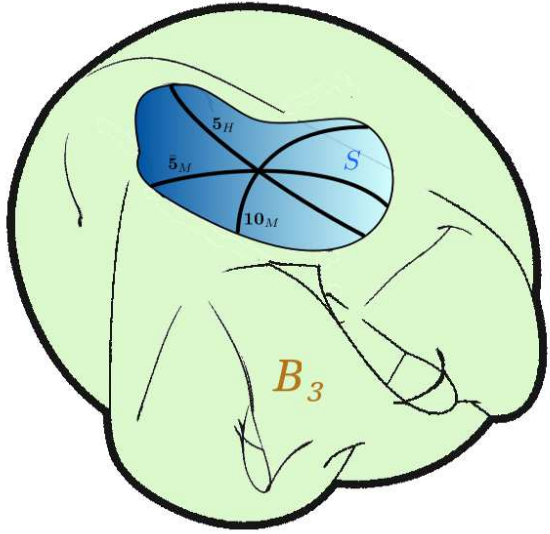}
\end{center}
\caption{Scheme of a IIB/F-theory SU(5) GUT.  The six extra dimensions are compactified on $B_3$ whereas the 
SU(5) degrees of freedom are localized on a  4-cycle submanifold $S$. The gauge bosons live 
on the bulk of $S$ but the chiral multiplets are localized on complex matter curves. At the intersection of
two matter curves with a Higgs curve a Yukawa coupling develops.}
\label{fthscheme}
\end{figure}

In a different development it has been realised that both gauge coupling unification 
as well as a large Yukawa coupling for the top quark can be obtained in a rather general form within the context of F-theory local SU(5) unification \cite{Donagi:2008ca, Beasley:2008dc, Beasley:2008kw, Donagi:2008kj, Heckman:2010bq, Weigand:2010wm, Maharana:2012tu,Leontaris:2012mh}. In this scheme the SU(5) 
degrees of freedom live on 7-branes wrapping a 4-dimensional manifold $S$ in the six extra
dimensions. The matter fields in the $\bar{\mathbf{5}}+\mathbf{10}$ of SU(5) are localized  on complex  1-dimensional \emph{matter curves} in $S$ (see figure \ref{fthscheme}). Chirality is due to the presence of magnetic open string 
fluxes $F_2$ within the curve and family replication comes from the degeneracy of zero modes of a Dirac equation in
the compact dimensions \cite{Bachas:1995ik,Angelantonj:2000hi,Blumenhagen:2000ea,Cremades:2004wa,Abe:2008fi}.

 One attractive aspect of this setting is that it is possible to write down 
explicit expressions for the internal wavefunctions of the physical zero modes,
at least on a local approximation of the equations of motion.  With these explicit expressions one can obtain
specific results for Yukawa couplings \cite{Heckman:2008qa, Hayashi:2009ge, Cecotti:2009zf, Hayashi:2009bt, Font:2009gq, Conlon:2009qq, Aparicio:2011jx, Font:2012wq} and certain higher-dimensional superpotential couplings like B-violating terms \cite{Camara:2011nj}.

%%%%%%%%%%%%%%%%%%%
%
%%%%%%%%%%%%%%%%%%%%%%%%%%%%

In this paper we combine for the first time these two ingredients (closed string fluxes and 
local wavefunctions for \emph{chiral} matter fields)\footnote{See also \cite{Camara:2009xy, Camara:2009zz} for previous related work on wavefunctions for vector-like matter fields in presence of closed string fluxes.} to address the issue of flavor changing soft terms. Although we 
concentrate on the general setting of type IIB orientifold/F-theory local SU(5)  compactifications, the
idea is quite general and can be applied to large classes of type IIA, type IIB and heterotic compactifications; and more generally, to
theories with extra dimensions in which chirality is induced by magnetic fluxes along the internal dimensions. In all such theories
with the gauge and matter degrees of freedom localized within a compact $2n$-dimensional manifold $S$ in the extra dimensions, SUSY-breaking masses arise from terms of the general form
\beq
\Phi_i(x)\Phi_j(x)^* \int _S\ d^nz\,d^n\bar z\ C(z,\bar z)\  \phi_i^{(0)}(z,\bar z) \phi_j^{(0)}(z,\bar z)^*
\label{general}
\eeq
where $z$ denote collectively the complex coordinates of $S$. The indices $i,j=1,2,3$ are SM generation indices and 
$\phi_i^{(0)}$ are zero mode wavefunctions corresponding to 4d SM matter fields $\Phi_i$. The function $C(z,\bar z)$ is a background 
factor,  that in the case of String Theory depends on  the internal geometry and  the closed and open string fluxes.  In general the internal wavefunctions of the three SM generations depend differently on the
compact dimensions. Taking an orthonormal basis for the zero modes,  
 the resulting 4d mass matrix is diagonal and degenerate when $C(z,\bar z)$ is constant over $S$. 
On the other hand, in the most general situation 
 on which $C(z,\bar z)$  is not constant, the 
mass matrix is generation dependent and non-diagonal.

In the context of local IIB/F-theory SU(5) models we can be more specific for the general
expression (\ref{general}). With that aim, in this work we compute the soft SUSY-breaking terms for matter fields localized on $\mathbf{5}$-plet matter curves by expanding the  Dirac-Born-Infeld and Chern-Simons (DBI+CS) 
7-brane action in the presence of closed string  ISD fluxes  $G_{(0,3)}$. This was already computed in ref.~\cite{Camara:2004jj} for
adjoint matter fields  living on the bulk of $S$. Here instead, by inserting the local  wavefunctions of MSSM  fields in the DBI+CS action, we generalise the above computation to the more interesting case 
of fields localized on matter curves, which is the relevant one for local F-theory SU(5) unification. 
Interestingly,  the soft terms that we obtain for constant flux density $G_{(0,3)}$ agree with those obtained from the effective $\mathcal{N}=1$ supergravity 
effective action for modulus dominance in \cite{Ibanez:2004iv,Lust:2004dn,Font:2004cx,Aparicio:2008wh}.
 Our  results however give  for the first time a microscopic and more model-independent derivation of the soft term structure, and allow to deal with more general cases than that of constant fluxes.

By allowing for {\it non-constant} local densities of closed and open string fluxes,
flavor violating soft masses do indeed generically appear.  We find that the main non-universal contribution of the flux densities leads at the unification scale to (see table \ref{tablafin} for more detailed results)
\beq
{\tilde \delta}_{12}  \ \simeq \ \frac { {\tilde m}_{12}^2}{ {\tilde m}^2} 
\ \simeq \ 
\left\{ \begin{array}{cl}    a \, (M_{\rm GUT}/M_{\rm Pl})^{1/3} & \quad \textrm{(cosed string flux)}\\ \nonumber
 b \,  (\alpha_{\rm GUT})^{1/2} & \quad \textrm{(open string flux)}
\end{array}\right.
\label{resuminitial}
\eeq
for the off-diagonal mixing of the first two generations. Here $\tilde m$ is the averaged sfermion mass and $a$ and $b$ are constants
of order one.  In the case of non-constant densities of closed string fluxes $G_{(0,3)}$ the induced off-diagonal mixing is parametrically suppressed by $(M_{\rm GUT}/M_{\rm Pl})^{1/3}$. 
This dependence can be easily understood: the smaller the volume of the 4-dimensional submanifold $S$ compared to the overall volume Vol$(B_3)\propto M_{\rm Pl}^2$, the less the 
flux density varies.  In the case of non-constant densities of open string fluxes, off-diagonal terms are proportional to
the flux density $F_2 \propto \alpha_{\rm GUT}^{1/2}$. 

Although parametrically suppressed,  the actual values $(M_{\rm GUT}/M_{\rm Pl})^{1/3}\simeq 0.1$ and $\alpha_{\rm GUT}^{1/2}\simeq 0.2$ are not that small and non-universalities at the unification scale are relatively large, of order $30 \ \%$. Despite of this, the renormalization group (RG) running of the
squark masses  down to low-energies dilutes substantially the mixing, giving rise to an extra suppression factor of order $\mathcal{O}(1/10)$
for squarks, so that the limits from $\Delta m_K$ in the kaon system are automatically 
satisfied for squark masses that are consistent with the current LHC limits. Hence, in the case of the $\tilde d$ and $\tilde s$ squarks the strongest bounds actually come from the CP violation
limits in kaons. The RG dilution is not enough in this case to get sufficient suppression and $\tilde d$ and $\tilde s$ squarks must have masses in the
 several TeV region. The bounds for sleptons are even tighter, as in this case the RG running to low-energies is smaller and does not
dilute the mixing sufficiently. In particular we find that to satisfy the bounds  coming from the yet unobserved $\mu\rightarrow e\gamma$ decay the sleptons are required to have masses in the 
$\gtrsim  10 $ TeV region, much above  LHC limits.       

Although there is considerable uncertainty
from model dependent factors 
(reflected in the  $a$ and $b$ coefficients in eq.~(\ref{resuminitial})), the general results seem to indicate that multi-TeV (or heavier) squarks and sleptons may be needed for consistency with experiment. Whereas this may look 
discouraging for the LHC, the presence of a Higgs mass around $m_H\simeq 126 $ GeV  already motivates models with heavy squark sectors. Our results point towards the same direction. On the other hand, as we describe in section 4, if the fluxes are not strongly varying (e.g. if a symmetry exists that forbids the leading
linear coordinate dependence)  the bounds are substantially relaxed and there is a chance of producing SUSY particles at LHC. 

The structure of this paper is as follows. In the next section we obtain general expressions for closed string flux induced soft terms
for the scalar fields that localize on matter curves of local IIB/F-theory SU(5) GUT's. The leading dependence of these soft terms on the
open string flux is also obtained. In section \ref{sec3} we compute the off-diagonal flavor violating mass terms induced by 
(non-constant) densities of closed and open string fluxes. We also discuss the constraints on squark and slepton masses
to have sufficiently suppressed FCNC processes, including CP violation in the kaon system as well as limits from $\mu \rightarrow e \gamma$.
  In section \ref{sec4} we briefly discuss under
what conditions the strong constraints on the squark/slepton masses from FCNC limits could be relaxed. The last section includes a
general discussion of the results and possible generalisations to other schemes.

\section{Soft terms for  local chiral   matter fields}
\label{sec2}

We consider type IIB compactifications with magnetized intersecting D7-branes (or their more general F-theory counterparts). We are in particular interested in compactifications that lead to local SU(5) GUT's  \cite{Donagi:2008ca,  Beasley:2008dc, Beasley:2008kw, Donagi:2008kj, Heckman:2010bq, Weigand:2010wm, Leontaris:2012mh,Maharana:2012tu}, although our results will apply to more general configurations. In those compactifications the SU(5) gauge degrees of freedom reside in the worldvolume of a single stack of D7-branes that wraps a 4-cycle $S$ in the compact 3-fold $B_3$. The 4d SU(5) gauge coupling is related to the size of $S$
(see e.g. \cite{Ibanez:2012zz})
\begin{equation}
\alpha_{\rm GUT}^{-1} = \frac{\textrm{Vol}(S)}{8\pi^4 g_s \alpha'^2}
\end{equation}
whereas gravitational effects are controlled by the overall volume of $B_3$, with the Planck's mass given by 
\begin{equation}
M_{\rm Pl} = \frac{\sqrt{2\textrm{Vol}(B_3)}}{4\pi^3 g_s\alpha'^2}
\end{equation}
Here $\alpha'$ and $g_s$ are the string slope  and coupling constant respectively. 
In terms of these the SU(5) unification scale reads
\begin{equation}
M_{\rm GUT}=  \left(\frac{2\alpha_{\rm GUT}}{\alpha'^2 g_s}\right)^{\frac14} \ .
\label{mgutdef}
\end{equation}

In what follows we assume that $B_3$ is such that Vol($B_3$) and Vol($S$) can be tuned independently. A remarkable property of such compactifications is that gravitational effects can be consistently decoupled from the SU(5) gauge dynamics by taking the volumes such that
\begin{equation}
\varrho^3\equiv \frac{\textrm{Vol}(S)^{3/4}}{\textrm{Vol}(B_3)^{1/2}}=\frac{\sqrt{2}}{\alpha_{\rm GUT}}\,\frac{M_{\rm GUT}}{M_{\rm Pl}}\ll 
1\label{decouple}
\end{equation}
In that case the 4d effective theory can be expanded in a power series of $\varrho$. In particular, for the observed values of $\alpha_{\rm GUT}^{-1}\simeq 24$ and $M_{\rm Pl}\simeq 1.2\times 10^{19}$ GeV this is a valid perturbative expansion as long as 
\begin{equation}
M_{\rm GUT} \ \ll \ 3.5 \times 10^{17}\ \textrm{GeV} \, .
\end{equation}

To leading order in this expansion, 7-branes are well described by 8d super Yang-Mills theory (SYM), twisted appropriately in order to account for the possibly non-trivial normal bundle \cite{Donagi:2008ca,  Beasley:2008dc}. For simplicity here we take a trivial normal bundle, so that the complex 4-cycle $S$ is actually Ricci-flat and the normal deformations of the D7-branes are parametrized in terms of a single 8d complex scalar field $\Phi$ that transforms in the adjoint representation of the gauge group.\footnote{The generalization to cases with non-trivial normal bundle is however straightforward, as it can be achieved by changing the magnetization along the canonical bundle in an amount proportional to the R-charge of the field (see e.g.~\cite{Conlon:2009qq}).} The Lagrangian for the 8d bosonic fields is therefore given by
\begin{equation}
\mathcal{L}_{\rm SYM}=\textrm{Tr}\left(D_a\Phi D^a\bar\Phi - \frac{1}{4}F_{ab}F^{ab}\right)
\label{lag}
\end{equation}
where $D_a$ are the gauge covariant derivatives.

Massless matter multiplets transforming in the $\mathbf{5}$ or $\mathbf{10}$ representations (or their conjugates) of SU(5) arise at the intersections of the GUT D7-branes with other 7-branes. Each intersection spans a complex curve or  \emph{matter curve} in $S$ along which the gauge symmetry is enhanced. The net 4d chirality of these multiplets is determined by the first Chern class of the magnetization along the matter curve and the number of intersections.

For concreteness let us consider the case of a $\bar{\mathbf{5}}$ matter curve localized at the intersection of the SU(5) stack with an additional U(1) D7-brane. Denoting by $x$ and $y$ the two local complex coordinates in $S$, we choose the intersection locus to be given by $x=0$. At this locus the SU(5) gauge symmetry is enhanced to SU(6) accordingly to the decomposition
\begin{align}
SU(6)&\, \to\, SU(5) \times U(1)\label{su6}\\
\mathbf{35}&\, \to\,  \mathbf{24_0}\, \oplus\, \mathbf{1_0}\, \oplus\, \mathbf{5_{-1}} \oplus \bar{\mathbf{5}}_{\mathbf{1}}\nonumber
\end{align}
The 4d effective theory can be thus described by dimensionally reducing (\ref{lag}) with SU(6) gauge group in the presence of a background for $\Phi$ of the form
\begin{equation}
\langle \Phi\rangle = \frac{m}{2}x Q\label{phiback}
\end{equation}
where $Q$ is the generator of the U(1) factor in eq.~(\ref{su6}), namely
\begin{equation}
Q=\frac{1}{\sqrt{60}}\, \textrm{diag}(1,1,1,1,1,-5)
\end{equation}
and $m$ is a parameter  (with mass$^2$ dimensions, see eq.~(\ref{laeta})) related to the intersection angle. We take the usual conventions for the normalization of the SU(6) generatores, $\textrm{Tr}(Q_i^2)=1/2$.

To have a chiral spectrum in 4d, we also consider the presence of a magnetization of the form
\begin{equation}
\langle F_2\rangle = iM(dx\wedge d\bar x -dy\wedge d\bar y)Q\label{magback}
\end{equation}
This particular structure for the magnetization satisfies the D-term equations and leads to a $\mathcal{N}=1$ supersymmetric spectrum in 4d
\cite{Font:2009gq,Cecotti:2009zf,Conlon:2009qq,Hayashi:2009bt,Aparicio:2011jx,Font:2012wq}.

Fields transforming in the adjoint representation of SU(6) can be decomposed accordingly to (\ref{su6}) as
\begin{equation}
\Phi = \left(\begin{tabular}{c|c}$\Phi_{\mathbf{24}}$ & $\Phi_{\mathbf{5}}$\\
\hline
$\Phi_{\bar{\mathbf{5}}}$ & $\Phi_{\mathbf{0}}$\end{tabular}\right)\ , \qquad A = \left(\begin{tabular}{c|c}$A_{\mathbf{24}}$ & $A_{\mathbf{5}}$\\
\hline
$A_{\bar{\mathbf{5}}}$ & $A_{\mathbf{0}}$\end{tabular}\right)\label{split}
\end{equation}
Upon dimensional reduction the 8d fields $\Phi_{\bar{\mathbf{5}}}$ and $A_{\bar{\mathbf{5}}}$ lead to three Kaluza-Klein (KK) towers of 4d complex scalars $\phi^{(k)}_i(x_\nu)$, $i=1,2,3$, and one KK tower of 4d gauge bosons $X^{(k)}_\mu(x_\nu)$,
\begin{align}
\Phi_{\bar{\mathbf{5}}}&=\sum_k \phi^{(k)}_3(x_\nu)\varphi ^{(k)} (z,\bar z)\label{kkphi}\\
A_{\bar{\mathbf{5}}}&= \sum_k\left[X_\mu^{(k)}(x_\nu)\chi^{(k)}(z,\bar z)dx^\mu\, +\, \left(\phi^{(k)}_1(x_\nu) a^{(k)}_{\bar x}(z,\bar z) d\bar x\, +\,  \phi^{(k)}_2(x_\nu) a^{(k)}_{\bar y}(z,\bar z) d\bar y\, +\, \textrm{c.c.}\right)\right]\nonumber
\end{align}
all of them transforming in the $\bar{\mathbf{5}}$ of SU(5). We make use of the short notations $(z, \bar z)$ and $(x_\nu)$ to represent a generic dependence on all the coordinates of $S$ and the 4d space-time directions respectively. In what follows we focus on 4d scalar zero modes. Their internal wavefunctions $a^{(0)}_{\bar x}$, $a^{(0)}_{\bar y}$ and $\varphi^{(0)}$ satisfy the following system of differential equations, derived from the equations of motion of (\ref{lag}),
\cite{Font:2009gq,Cecotti:2009zf,Conlon:2009qq,Hayashi:2009bt,Aparicio:2011jx,Font:2012wq}
\begin{equation}
\mathbb{M}\Psi^{(0)}_{\bar{\mathbf{5}}}(z,\bar z) = 0
\end{equation}
where
\begin{equation}
\mathbb{M}=\begin{pmatrix}\triangle+ qM & 0 & \frac{qm}{2}\\
0 & \triangle-qM & 0\\
\frac{qm}{2} & 0 & \triangle\end{pmatrix}\ , \qquad \Psi^{(0)}_{\bar{\mathbf{5}}}(z,\bar z)=\begin{pmatrix}a^{(0)}_{\bar x}\\ a^{(0)}_{\bar y}\\ \varphi^{(0)} \end{pmatrix} \ , \qquad \triangle = -\sum_{j=1}^3 D_j^\dagger D_j
\end{equation}
and
\begin{align}
D_1 &= \partial_x - \frac{qM}{2}\bar x & D_1^\dagger &= \bar\partial_{\bar x}+\frac{qM}{2}x\\
D_2 &= \partial_y + \frac{qM}{2}\bar y & D_2^\dagger &= \bar\partial_{\bar y}-\frac{qM}{2}y\nonumber\\
D_3 &= -\frac{qm}{2}\bar x & D_3^\dagger &=\frac{qm}{2} x\nonumber
\end{align}
with $q=\sqrt{\frac35}$ the U(1) charge normalisation of $\Psi^{(0)}$. The reader may easily check that for $M < 0$ there is a set of $r$ massless scalars with normalizable wavefunction given by 
\begin{equation}
\Psi^{(0)}_{\bar{\mathbf{5}}_i}(z,\bar z)=\frac{1}{\sqrt{2\lambda(\lambda-M)}}\begin{pmatrix}M-\lambda\\ 0\\ m\end{pmatrix} \psi^+_i(z,\bar z) \ , \qquad i=1,\ldots, r
\label{wave1}
\end{equation}
where we have defined the functions
\begin{equation}
\psi^\pm_i(z,\bar z)\equiv N_if_i(y)\, \textrm{exp}\left[-\frac{q\lambda}{2}|x|^2 \pm \frac{qM}{2}|y|^2\right]\label{psipm}
\end{equation}
Here $N_i$ is a normalization constant, $\lambda \equiv \sqrt{M^2+m^2}$ and $f_i(y)$ are holomorphic functions of $y$. The degeneracy $r$ and the holomorphic functions $f_i(y)$ can only be determined in terms of the global topology of $S$ and the gauge bundle.

Similarly, for $M>0$ there is a set of $r$ massless scalars transforming in the $\mathbf{5}$ of SU(5) with normalizable wavefunction given by 
\begin{equation}
\left.\Psi^{(0)}_{\mathbf{5}_i}(z,\bar z)\right.^\dagger=\frac{1}{\sqrt{2\lambda(\lambda+M)}}\begin{pmatrix}M+\lambda\\ 0\\ m\end{pmatrix} \psi^-_i(z,\bar z) \ , 
\qquad i=1,\ldots, r   \ \ .
\label{wave2}
\end{equation}
For $M\neq 0$ the wavefunctions $\Psi^{(0)}_{\mathbf{5}_i}$ and $\Psi^{(0)}_{\bar{\mathbf{5}}_i}$ are mutually exclusive, being only one of the two sets normalizable, and corresponding to a chiral spectrum in 4d. Since the local background that we have taken preserves $\mathcal{N}=1$ supersymmetry, the above set of massless scalars are actually the bosonic components of $r$ massless $\mathcal{N}=1$ chiral supermultiplets.

In a complete model the right-handed down squarks and the left-handed sleptons are contained in the $\bar{\mathbf{5}}$ of SU(5) and therefore are well described by the above setup with $M<0$.\footnote{Alternatively, in the case where $F_2$ also vanishes, the above matter curve would instead lead to a set of $r$ vector-like pairs of $\mathcal{N}=1$ supermultiplets transforming in the $\mathbf{5} \oplus \bar{\mathbf{5}}$ of SU(5). In a complete model these would be identified with $r$ Higgs-up and $r$ Higgs-down multiplets.}{}$^{,}$\footnote{In this work we only consider models where all three families arise from the same matter curve. Alternatively different families could arise from different matter curves \cite{Dudas:2009hu,Leontaris:2010zd}. Although a detailed analysis is required, in that case we expect larger FCNC's because wavefunctions of different families are typically localized at points in $S$ that are farther away from each other.} Moreover we take $r=3$, corresponding to the three families of the Standard Model. In that case we can take a basis such that the expansion of the holomorphic functions $f_i(y)$ around the origin reads
\begin{equation}
f_i(y)=y^{3-i} + \mathcal{O}\left(y^4\right)\ , \qquad i=1,2,3 \label{holofact}
\end{equation}
In particular wavefunctions of different families localize in slightly different regions of the 4-cycle $S$ due to their different holomorphic pieces $f_i(y)$. 

Due to the Gaussian localization in both $x$ and $y$ the replacement $S\sim \mathbb{C}^2$ is a suitable approximation when Vol($S$) is large in string units. Normalizing the wavefunctions in $\mathbb{C}^2$ we get
\begin{equation}
N_i=\frac{\sqrt{q\lambda} |qM|^{\frac{4-i}{2}}}{\pi\sqrt{(3-i)!}}\ .
\end{equation}

Having the wavefunctions of the 4d chiral fields in a SU(5) GUT model one can compute their couplings in terms of integrals of overlaps of wavefunctions, as dictated by dimensional reduction of the 8d Lagrangian (\ref{lag}). Since the wavefunctions are localized in the internal dimensions, the integrals receive only contributions from the background in a local patch around the localization point. Such approach has in particular turned out to be very fruitful in studying the flavor structure of Yukawa couplings \cite{Heckman:2008qa, Hayashi:2009ge, Cecotti:2009zf, Hayashi:2009bt, Font:2009gq, Conlon:2009qq, Aparicio:2011jx, Font:2012wq}, where it has been shown that in the limit $\varrho \to 0$ only the Yukawa coupling of the third family is non-zero, providing a natural origin for the hierarchical mass of the top quark in the Standard Model.

In what follows we would like to extend the above techniques to the computation of soft SUSY-breaking terms. Note that the background  near the localization point, given in eqs.~(\ref{phiback})-(\ref{magback}), preserves $\mathcal{N}=1$ supersymmetry in 4d and therefore soft terms for the GUT multiplets vanish. 
Hence, we should first extend the local background to a more general one that breaks supersymmetry, for instance through the presence of closed string fluxes. In a full-fledged global setup, such local density of supersymmetry-breaking flux may arise from the presence of topological fluxes along some of the cycles of $B_3$ but also from the backreaction of distant localized sources that break supersymmetry, such as e.g distant anti-branes or non-perturbative effects leading to dynamical SUSY-breaking. 
Whereas closed string fields are non-dynamical in the limit (\ref{decouple}), the 4d soft terms induced by their backgrounds survive to that limit (recall that they are defined in the limit $M_{\rm Pl}\to \infty$) as long as the gravitino mass $m_{3/2}$ is kept finite.

For simplicity here we only consider a background near the GUT-branes given by a local density of RR and NSNS 3-form fluxes $G_3=F_3-\tau H_3$ of type $(0,3)$ with respect to the local complex structure, where $\tau = C_0 + i e^{-\phi}$ is the complex axio-dilaton. Such local configuration of the background arises naturally
e.g.  in the Large Volume Scenario \cite{Balasubramanian:2005zx, Conlon:2005ki, Conlon:2006wz, Conlon:2007xv, Blumenhagen:2009gk}, where the dominant source of SUSY-breaking is the F-term of the K\"ahler modulus that controls the size of the 4-cycle $S$. Soft-terms for magnetized intersecting D7-branes probing more general flux backgrounds will be considered in \cite{inprogress}.

In the presence of a non-trivial density of 3-form flux near the stack of 7-branes the 8d Lagrangian (\ref{lag}) receives extra contributions from the flux. The leading contributions for D7-branes were computed in \cite{Camara:2004jj} by expanding the non-Abelian DBI+CS action \cite{Myers:1999ps} in powers of the transverse fluctuations $\Phi$, from which a microscopic computation of the flux-induced soft terms for non-magnetized non-intersecting D7-branes was performed. Here we are interested in doing an analogous microscopic computation for intersecting magnetized D7-branes. For our current purposes the relevant piece of the D7-brane DBI+CS action is
\begin{equation}
S_{\rm 8d}=\mu_7 g_s \textrm{Tr}\left[-\int_{\mathbb{R}^{1,3}\times S} d^8\xi \sqrt{-\textrm{det}\left(g_{ab}+8\pi^2\alpha'^2 D_{(a}\Phi D_{b)}\bar\Phi - g_s^{-1/2}\mathcal{F}_{ab}\right)}+\int_{\mathbb{R}^{1,3}\times S} C_8\right]
\end{equation}
where we are working in the 10d Einstein frame, $\mu_7=(2\pi)^{-7}\alpha'^{-4}g_s^{-1}$ is the D7-brane tension and we have defined 
\begin{equation}
\mathcal{F}_2\equiv B_2-2\pi\alpha' F_2
\end{equation}
Expanding the determinant and the square root to quadratic order in the fields we get
\begin{equation}
\mathcal{L}_{\rm 8d}=\mu_7 g_s\, \textrm{Tr}\left(4\pi^2\alpha'^2 D_a\Phi D_a\bar\Phi-\frac{g_s^{-1}}{4}\mathcal{F}_{ab}\mathcal{F}_{ab}+C_8\right)\label{interl}
\end{equation}
In the limit (\ref{decouple}) the dilaton, the B-field and the RR 8-form become auxiliary fields tied to the transverse fluctuations of the brane. We take
a gauge in which
\begin{equation}
B_2 = g_s\alpha'\pi i\left(G(z,\bar z)\,\bar\Phi\, d\bar x\wedge d\bar y - G^*(z,\bar z)\,\Phi\, dx\wedge dy\right) + \ldots
\end{equation}
with $G(z,\bar z)$ a complex function of the coordinates that parametrizes the density of (0,3)-flux in $S$. The dots in this expression represent higher powers of the transverse fluctuations. With this gauge choice the pull-back of the B-field to the D7-brane worldvolume becomes trivial, $P[B_2]=B_2$. Making use of the 10d type IIB supergravity equations of motion it is possible to show that the RR 8-form potential in this background is related to $B_2$ as \cite{Camara:2004jj,inprogress}
\begin{equation}
C_8=-\frac{g_s^{-1}}{2}B_2\wedge B_2\wedge  dx_0\wedge dx_1\wedge dx_2\wedge dx_3 
\end{equation}
where $x_\nu$, $\nu=0,\ldots,4$, are the coordinates along the non-compact directions. Hence, eq.~(\ref{interl}) finally becomes 
\begin{equation}
\mathcal{L}_{\rm 8d}=\textrm{Tr}\left(D_a\Phi D^a\bar\Phi\, -\, \frac{1}{4} F_{ab}F^{ab}\, -\, \frac{g_s}{2}|G|^2|\Phi|^2\, +\, \ldots \right)\label{lagmas}
\end{equation}
where we have rescaled the fields to have canonically normalized kinetic terms. Observe that the local density of 3-form flux induces a 8d mass term for $\Phi$. Upon dimensional reduction in the presence of non-trivial backgrounds $\langle \Phi \rangle$ and $\langle F_2\rangle$ this leads to 4d soft masses for the fields localized at the matter curves, as we show now.

Indeed, let us consider again the above example of the $\bar{\mathbf{5}}$ matter curve containing the right-handed down squarks and the left-handed sleptons of a local SU(5) GUT and include a non-trivial density of 3-form flux $G$ near the origin of coordinates where the wavefunction (\ref{wave1}) localizes. We saw respectively in (\ref{wave1}) and (\ref{wave2}) that fields transforming in the $\bar{\mathbf{5}}$ (normalizable for $M<0$) and in the $\mathbf{5}$ (normalizable for $M>0$) of SU(5) originate from a mixture between $A_{\bar x}$ and $\Phi$. More precisely,
\begin{equation}
\begin{pmatrix}\phi^{(0)}_{\bar{\mathbf{5}}_i}(x_\nu) \psi^+_i(z,\bar z)\\ \phi^{(0)}_{\mathbf{5}_i}{}^\dagger(x_\nu)\psi^-_i(z,\bar z)\end{pmatrix}=\begin{pmatrix}\frac{M-\lambda}{\sqrt{2\lambda(\lambda-M)}} & \frac{m}{\sqrt{2\lambda(\lambda-M)}}\\
\frac{M+\lambda}{\sqrt{2\lambda(\lambda+M)}} & \frac{m}{\sqrt{2\lambda(\lambda+M)}}
 \end{pmatrix}\begin{pmatrix}A^{(0)}_{\bar{\mathbf{5}}_i,\bar x}\\ \Phi^{(0)}_{\bar{\mathbf{5}}_i}\end{pmatrix}\label{so2}
\end{equation}
where $\phi^{(0)}_{\bar{\mathbf{5}}_i}(x_\nu)$ and $\phi^{(0)}_{\mathbf{5}_i}(x_\nu)$ are respectively the 4d scalars in the $\bar{\mathbf{5}}_i$ and the $\mathbf{5}_i$ of SU(5), and $\psi^\pm_i(z,\bar z)$ were defined in eq.~(\ref{psipm}). This relation can be inverted in order to express $\Phi^{(0)}_{\bar{\mathbf{5}}_i}$ as
\begin{equation}
\Phi^{(0)}_{\bar{\mathbf{5}}_i}\, =\, \sqrt{\frac{\lambda-M}{2\lambda}}\frac{\lambda+M}{m}\, \phi^{(0)}_{\bar{\mathbf{5}}_i}(x_\nu) \psi^+_i(z,\bar z)\, +\, \sqrt{\frac{\lambda+M}{2\lambda}}\frac{\lambda-M}{m}\, \phi^{(0)}_{\mathbf{5}_i}{}^\dagger(x_\nu)\psi^-_i(z,\bar z)
\end{equation}
and therefore the 8d mass term for the corresponding block of $\Phi$ (c.f (\ref{split})) decomposes in internal and external parts as
\begin{equation}
|\Phi_{\bar{\mathbf{5}}}|^2 \, =\,  \left|\sum_k\sum_{i=1}^3 \Phi^{(k)}_{\bar{\mathbf{5}}_i}\right|^2\, =\, \frac12\sum_{i,j=1}^3\left[1 - \left|\frac{M}{m}\right| + \mathcal{O}\left(\left|\frac{M}{m}\right|^3\right)\right]\phi_{\bar{\mathbf{5}}_i}^{(0)}\phi_{\bar{\mathbf{5}}_j}^{(0)}{}^\dagger \psi_i^+(\psi_j^+)^*\, +\, \ldots\label{mass8d}
\end{equation}
where we have expanded perturbatively in powers of the ratio $|M/m|$ between the magnetization and the intersection parameters. Higher-order terms in the brackets contain only odd powers of $|M/m|$. 

Since (\ref{so2}) is a SO(2) transformation, the 8d kinetic terms are simply reduced as
\begin{equation}
D_\mu \Phi_{\bar{\mathbf{5}}} D^\mu \Phi_{\bar{\mathbf{5}}}^\dagger\, +\, D_\mu A_{\bar{\mathbf{5}},\bar 1}D^\mu A_{\bar{\mathbf{5}},\bar 1}^\dagger\, =\, \sum_{i,j=1}^3\psi_i^+(\psi_j^+)^*\, D_\mu \phi_{\bar{\mathbf{5}}_i}^{(0)} D^\mu \phi_{\bar{\mathbf{5}}_i}^{(0)}{}^\dagger \, + \, \ldots\label{kinet8d}  
\end{equation}
Plugging (\ref{mass8d}) and (\ref{kinet8d}) into (\ref{lagmas}) and integrating over the 4-cycle $S$, we obtain the expression for the 4d sfermion soft mass matrix of the $\bar{\mathbf{5}}$ matter curve.\footnote{One may worry about possible corrections from the closed string fluxes to the wavefunctions (\ref{wave1}) and (\ref{wave2}), since the latter were derived in absence of closed string fluxes. Those corrections indeed arise (see \cite{Camara:2009xy, inprogress}). However, for our current purposes they can be safely ignored, since in this setting the scale of SUSY-breaking is much smaller than the unification and string scales, $M_{\rm SS}\ll M_{\rm GUT},\, M_{\rm st}$.} To linear order in the magnetization it is given by
\begin{equation}
\boxed{m_{ij}^2 = \frac{g_s}{4\textrm{Vol}(S)}\int_S d^2zd^2\bar z \sqrt{g_4}\, |G|^2\left(1-\left|\frac{M}{m}\right|\right)\psi_i^+(\psi_j^+)^*}\label{massfinal}
\end{equation}
where $g_4$ is the determinant of the metric in $S$ and 
$\textrm{Vol}(S)=\int_S d^2zd^2\bar z \sqrt{g_4}$. Kinetic terms are diagonal and canonically normalized.

Equation (\ref{massfinal}) is the main result of this section. For constant densities of magnetization $M(z,\bar z)=M_0$ and 3-form flux $G(z,\bar z)=G_0$ the integral becomes trivial and the standard supergravity formula for the soft scalar masses in intersecting magnetized 7-branes is recovered~\cite{Ibanez:2004iv,Lust:2004dn, Font:2004cx,Aparicio:2008wh}
\begin{equation}
m_{ij}^2 = \frac{g_s}{4}\delta_{ij} |G_0|^2\left(1-\left|\frac{M_0}{m}\right|\right)\label{universal} \ .
\end{equation}
This mass matrix is diagonal and flavor universal, so that there are not new FCNC's induced beyond those of the Standard Model.
The correction linear in the fluxes was already advanced in refs.\cite{Font:2004cx,Aparicio:2008wh}. In this constant case one also finds in the absence of open string fluxes that $M_{1/2}^2=g_s|G_0|^2/2$.  
 From the point of view of the 4d effective supergravity these soft terms arise 
 from a non-vanishing F-term of the K\"ahler modulus that controls the size of the GUT 4-cycle and is therefore proportional, 
to leading order, 
to the universal gaugino mass $M_{1/2}$ 
\begin{equation}
m^2_{ij} = \frac{M^2_{1/2}}{2}\delta_{ij} \ .
\end{equation}

\begin{figure}[!ht]
\begin{center}
\includegraphics[scale=.66]{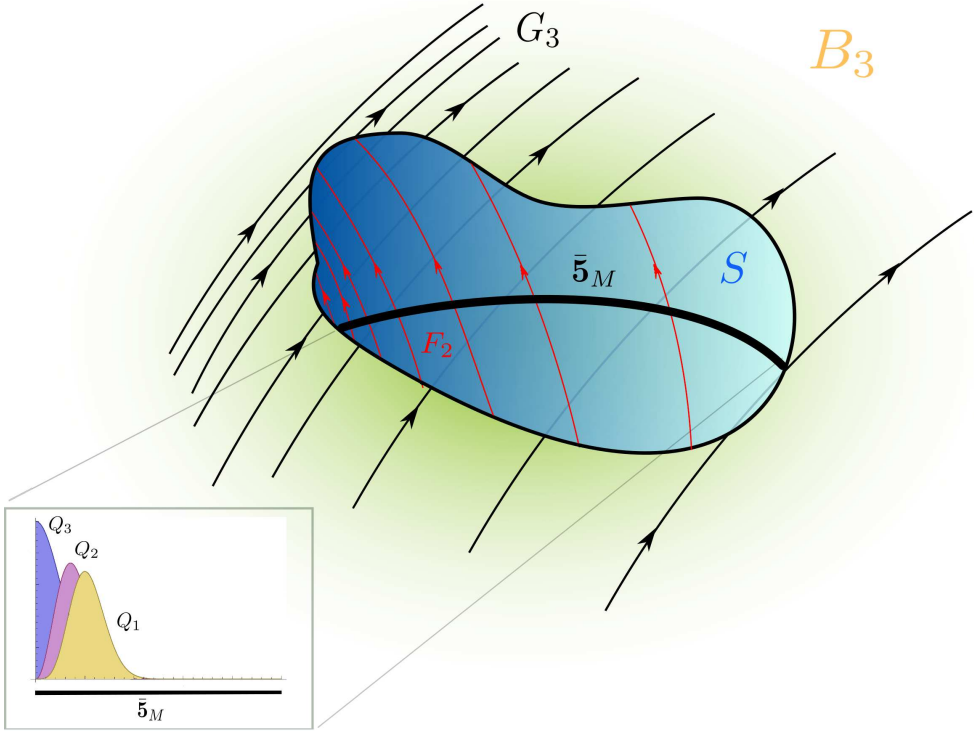}
\vspace{0.3cm}
\caption{\small The 4-cycle $S$ embedded in the ambient 3-fold $B_3$. The density of three-form flux $G_3$ (black arrows) and magnetization $F_2$ (red arrows) vary over $S$. Since wavefunctions of different families are localized in different regions of the $\bar{\mathbf{5}}_M$ matter curve, each family feels a different density of $G_3$ and $F_2$, leading to non-universal soft masses in the 4-dimensional effective theory.\label{figura}}
\end{center}
\end{figure}

However, in general there is no reason for the flux densities $M(z,\bar z)$ and $G(z, \bar z)$ to be constant over the GUT 4-cycle $S$. In that case non-universal soft masses are expected to arise from eq.~(\ref{massfinal}). Morally, as different families of sfermions are localized in slightly different regions of $S$ because of their holomorphic factors (\ref{holofact}), each family feels a different density of 3-form flux and/or magnetization and therefore gets different soft masses, see figure \ref{figura}.

%%%%%%%%%%%
%%%%%%%%%%%

As we have already commented, such (non-constant) flux densities can arise from the backreaction of distant sources or non-perturbative effects. The local background near the GUT D7-branes can be understood in that case as originating from massive open string modes that propagate between the distant sources and the stack of GUT branes, and eq.~(\ref{massfinal}) computes the thresholds to the sfermion mass matrix that result from integrating out those heavy modes, in the same spirit than the supergravity computation of gauge thresholds \cite{Giddings:2005ff, Baumann:2006th}.

In the remaining part of the paper we make use of eq.~(\ref{massfinal}) to estimate the size of non-universal soft masses in local SU(5) GUT's with non-constant magnetization and 3-form flux and compare the resulting FCNC transitions  with the current experimental bounds.

%%%%%%%%%%%%%%%%%%%%%%%%%%%%%%%%%%%%%%%%%%%%%%%%%%%%%%

\section{Flavor non-diagonal soft terms}
\label{sec3}

%%%%%%%%%%%%%%%%%%%%%%%%%%%%%%%%%%%%%%%%%%%%%%%%%%%%%%

In the leptonic sector, the strongest flavor violating constraints come from limits for  the branching ratio
for $\mu\rightarrow e \gamma$. 
From the hadronic constraints, the strongest limits on flavor dependent soft terms come from the kaon system,
in particular from $K^0-\overline{K}^0$ oscillations and CP violation. The relevant quantities are the real and imaginary parts of the off-diagonal 
$\tilde d-\tilde s$ squark and $\tilde e - \tilde \mu$ slepton soft masses ${\tilde m}_{12}^2$.  In both the leptonic and hadronic sectors,
to directly compare with the experimental 
constraints it is customary to work in a fermion basis in which the   fermion mass matrix is diagonal.
In particular, for the kaon system one has
\begin{equation}
{\tilde m}_{ij}^2\ =\left(U_{\rm d}\, m_{\rm soft}^2\, U_{\rm d}^\dag \right)_{ij} \ ,\qquad i,j=1,2,3
\label{defm12}
\end{equation} 
where $m_{\rm soft}^2$ is the original squark mass matrix before quark diagonalisation and $U_{\rm d}$ is the unitary matrix that diagonalises the down-quark mass matrix.
It is also customary to work in the mass insertion approximation \cite{Hagelin:1992tc} where we expand on the ratio of the off-diagonal terms over the averaged squark mass
${\tilde m}_{\tilde q}^2$,
\beq
{\tilde \delta}_{ij}^{\rm d} \ =\frac {{\tilde m}^2_{ij}}{{\tilde m}_{\tilde q}^2} \ .
\eeq

There are in fact four $3\times 3$ squark mass matrices (\ref{defm12}), depending on the particular \emph{chirality} of the squarks,
namely $m_{\rm RR}^2,\, m_{\rm LL}^2,\, m_{\rm LR}^2$ and $m_{\rm RL}^2$.  In the scheme for local SU(5) GUT's that we have described in the previous section, $m_{\rm LR}^2$ and $m_{\rm RL}^2$ are much suppressed for the two lightest generations, which are the ones relevant for the kaon system. This is because 
left- and right-handed squarks live in different matter curves and their mass terms are proportional to the 
Yukawa couplings, being those negligible for the first two generations. On the other hand 
$m_{\rm LL}^2$ and $m_{\rm RR}^2$ have both an analogous structure so that our results  below apply to
$m_{\rm RR}^2$ or $m_{\rm LL}^2$ irrespectively. Note that the down-quark mass matrix is in general not
symmetric and it  is actually diagonalized by
a bi-unitary transformation involving matrices $U_{\rm d}^{\rm R}$ and $U_{\rm d}^{\rm L}$ simultaneously. Thus in eq.~(\ref{defm12}) we should actually take $U_{\rm d}=U_{\rm d}^{\rm L}$ for 
${m}^2_{\rm LL}$ and $U_{\rm d}=U_{\rm d}^{\rm R}$ for ${m}^2_{\rm RR}$.

Focusing on the first two generations and taking $U_{\rm d}$ real for simplicity, we can parametrize $U_{\rm d}$ by an orthogonal 
$2\times 2$ matrix 
\beq 
U_{\rm d}\, =\, \begin{pmatrix}
  \cos\theta &   \sin\theta\\   
  -\sin\theta & \cos\theta \\
\end{pmatrix}
\eeq
From eq.~(\ref{defm12}) we obtain
\begin{equation}
{\tilde \delta}_{12}^{\rm d} \, =\, \frac { 2m_{12}^2 \cos 2\theta  +  (m_{22}^2-m_{11}^2) \sin 2\theta }{2\tilde m_{\tilde q}^2} \label{d12til}\ .
\end{equation}
Barring accidental cancellations, the value of ${\tilde \delta}_{12}^d$ is therefore controlled by 
the ratios $\delta_{12}^{\rm d}\equiv m_{12}^2/{\tilde m^2_{\tilde q}}$ and  $\rho_{12}^{\rm d}\equiv(m_{22}^2-m_{11}^2)/{2\tilde m^2_{\tilde q}}$. 
Totally analogous formulae, with the obvious changes, may be written for the slepton mass matrices relevant
for the $\mu \rightarrow e \gamma$ decay. 

In the next two subsections we evaluate the size of $\tilde\delta_{12}^{}$ 
 in local SU(5) GUTs by making use of the expression (\ref{massfinal}) for the soft scalar masses that we have derived microscopically in the previous section. Note however that such expression is only valid at the unification scale $M_{\rm GUT}$. To compare 
with the low-energy data we therefore have to take into account the renormalisation group (RG) running from the unification scale down to the 
electroweak scale. Since the Yukawa couplings of the first two generations are negligible, it is only the  SUSY-gauge couplings that contribute to this
running. Integrating the RG equations we find \cite{Ibanez:1984vq}
\begin{equation}
m^2(1^{\rm st},2^{\rm nd} \text{gen.})=m_0^2+g(t)M_{1/2}^2
\label{mrun}
\end{equation}
where
we are assuming universal gaugino  and scalar masses, $M_{1/2}$ and $m_0$, at the scale $M_{\rm GUT}$.\footnote{The effect of the non-universalities of the squark masses on the running can be safely neglected.}
We have defined
\begin{equation}
g(t)=2\sum_{k=1}^3C_kb_k\left(1-\frac{1}{(1+\beta_k t)^2}\right)
\end{equation}
where $C_k$ is the quadratic Casimir corresponding to the particular scalar field (namely, $C_k=\frac{N^2-1}{2N}$, $k=2,3$, for the fundamental of SU$(N)_k$ and $C_1=Y^2$ for U(1)${}_Y$) and the $\beta$-functions are
\begin{equation}
\beta_k=\frac{\alpha_k(0)b_k}{4\pi} \ .
\end{equation}
For the evaluation of the $\beta$-functions we consider a MSSM spectrum, which yields
 $(b_1,b_2,b_3)=(11,1,-3)$ with  $\alpha_3(0)=\alpha_2(0)=\frac{5}{3}\alpha_1(0)=\alpha_{\rm GUT}\simeq \frac{1}{24}$  the gauge unification constants.
Finally,  $t=2\log{(M_{\rm GUT}/M_{\rm SS})}$ where $M_{\rm SS}$ is the scale of supersymmetry-breaking. 

In the case of squarks, the leading contribution to the running (\ref{mrun}) comes from its SUSY-QCD part and thus the averaged squark mass ${\tilde m}_{\tilde q}$ has a substantial low-energy running. On the other hand, due to the 
 universality of  gauge couplings both  $(m_{22}^2-m_{11}^2)$ 
and $m_{12}^2$ have negligible 
low-energy running.  This is important because it means that in going from $M_{\rm GUT}$ down to
$M_{\rm SS}$, the ratio ${\tilde \delta}_{12} $ is diluted by a RG factor which is typically of order
\beq
{\tilde \delta}_{12}^{\rm d} \ \to  \   \frac {{\tilde \delta}_{12}^{\rm d}}  {1+g(t)\xi^2 }\ ,\qquad  \xi\equiv\frac {M_{1/2}}{m_0} \ .
\label{d12}
\eeq
For instance, for the type of 3-form fluxes with $M^2_{1/2}=2m^2_0$ that we have discussed in the previous section, and taking  
$M_{\rm GUT}\simeq 10^{16}$ GeV and $M_{\rm SS}\simeq 2 $ TeV, we obtain  $g(t)\simeq 4.5 $ and therefore a suppression  of order
$\simeq 1/10$ for the ratio ${\tilde \delta}_{11}^{\rm d}$ with respect to its value at the unification scale. 
This {\it RG dilution} turns out to be important in comparing with the low-energy data and, in particular, in evaluating the squark mass limits.
On the other hand for the case of sleptons the dilution is weaker and for the above parameters $g(t)\simeq 0.5$ so the suppression is only
of order $\simeq 1/2$. This different dilution will eventually make the slepton limits stronger than those coming from the kaon system.

In what follows we compute the flavor mixing induced by variations of closed and open string fluxes. We consider these two cases separately since, as it will become clear below,  terms coming from the simultaneous variation 
of closed and open string fluxes are either suppressed or contribute to transitions between the first and third families rather
than between the two lightest families.

\subsection{Flavor mixing  from non-constant closed string fluxes}

For concreteness we consider the $\bar{\mathbf{5}}$ matter curve $x=0$ that we have introduced in section \ref{sec2}, with a non-constant density of closed string 3-form flux $G(z,\bar z)$ and a constant density of magnetization $M_0$ along the 4-cycle $S$. The structure for the local background is as described in section \ref{sec2}. The soft scalar masses for the right-handed down squarks and the left-handed sleptons that are contained in the $\bar{\mathbf{5}}$ matter curve can be computed microscopically by means of eq.~(\ref{massfinal}). Due to the Gaussian localization of the wavefunctions $\psi_i^+$ around the point $x=y=0$, the dominant contribution to the integral in (\ref{massfinal}) comes from the background near the localization point. It is therefore convenient to perform an expansion of the closed string flux density in powers of the local coordinates $x$, $y$ of the 4-cycle
\beq
|G(z,\bar z)|^2\, =\, |G_0|^2 \left(1\, +\,  G_{y}^*\,y+G_{y}\,\bar y  \, +\,   G_{y\bar y}  \,  |y|^2 \, +\, \ldots \right)
\eeq
where $G_0$, $G_{y}$, here defined,  are complex constants and $G_{y\bar y}$ is real.
We have only displayed terms of the expansion that contribute to the flavor dependence of
the two lightest families. In particular, we have not shown the expansion in $x$ since it has no consequences for the  flavor dependence. 

For sufficiently large sizes of $S$ we can extend the domain of integration to infinity, so that (\ref{massfinal}) reads
\begin{multline}
m_{ij}^2 = \frac{g_s N_iN_j}{4}\int_{\mathbb{C}^2}  d^2x d^2y  \, \left[ |\hat G_0|^2 \left(1\, +\,  G_{y}^*y+G_{y}\bar y  \, +\,   G_{y\bar y} |y|^2 \, +\, \ldots \right)\right.\\
\left.y^{3-i}\bar{y}^{3-j}\, e^{-q\lambda|x|^2-q|M_0||y|^2}\right]
\label{m2}
\end{multline}
where we have defined
\begin{equation}
|\hat G_0|^2\equiv |G_0|^2\left(1-\left|\frac{M_0}{m}\right|\right) 
\end{equation}

Let us compute first the diagonal terms $i=j$. Linear terms  on $x,\, y$ vanish upon integration, so that the only non-vanishing contributions are
\begin{multline}
m_{ii}^2= \frac{g_s N_iN_j}{4}\int_0^\infty 2\pi x\, dx\int_0^\infty 2\pi y\, dy    \,
|\hat G_0|^2\left(1+G_{y\bar y}
|y|^2\right)e^{-q\lambda|x|^2-q|M_0||y|^2}|y|^{2(3-i)}\\ 
= \frac{g_s}{4}|\hat G_0|^2\left(1+G_{y\bar y}\frac{4-i}{q|M_0|}\right)  \ .
\label{mii}
\end{multline}
Similarly, for the off-diagonal $\Delta F = 1$ soft masses we have
\begin{multline}
m_{ij}^2= \frac{g_s N_iN_j}{4}\int_0^\infty 2\pi x\, dx\int_0^\infty 2\pi y\, dy \, |\hat G_0|^2 \left(G_y^*y  + G_y\bar y\right) e^{-q\lambda|x|^2-q|M_0||y|^2}y^{3-i}\bar{y}^{3-j}\\
=\frac{g_sk}{4}\frac{|\hat G_0|^2G_y}{\sqrt{q|M_0|}}\ , \qquad \textrm{where} \qquad k\equiv\left\{\begin{tabular}{c}$\sqrt{2}\quad$ for $\quad i=1,\ j=2$\\
$1\quad$ for $\quad i=2,\ j=3$
\end{tabular}\right.
\label{mij}
\end{multline}
Finally, 
the off-diagonal $\Delta F = 2$ mass term $m_{13}^2$  is proportional to higher derivatives of the 3-form flux and is therefore  subleading with respect to $m_{12}^2$ and $m_{23}^2$.
 
Summing  up, the structure of the soft mass matrix (\ref{m2}) is given by
\begin{equation}
\begin{pmatrix}m_{\tilde q}^2+\delta m^2_{1}&m_{12}^2&m_{13}^2\\
(m_{12}^{2})^*&m_{\tilde q}^2+\delta m^2_{2}&m_{23}^2\\
(m_{13}^{2})^*&(m_{23}^{2})^*&m_{\tilde q}^2+\delta m^2_{3}
\end{pmatrix}
\label{matrix}
\end{equation}
where $m_{\tilde q}^2$ is the universal soft mass for constant density of fluxes, eq.~(\ref{universal}), and the flavor violating terms have the following hierarchical structure
\begin{equation}
m_{\tilde q}^2\ >\ m_{12}^2,\, m_{23}^2\ >\ \delta m_i^2,\, m_{13}^2
\end{equation}
Making use of eqs.~(\ref{mii}) and (\ref{mij}) we can estimate the size of the non-universalities for the  first two families in a generic model. Indeed, flux quantization gives us an estimate  of the dependence of the flux densities on the volumes of $S$ and $B_3$ 
\begin{align}
\int_{\Sigma_i}{F_2}&=2\pi n_i\quad \Rightarrow \quad M_0\sim \frac{2\pi n}{\textrm{Vol}(S)^{1/2}}
\label{F}\\
\frac{1}{2\pi \alpha'}\int_{\gamma_j}G_3&= 2\pi f_j\quad \Rightarrow \quad G_0\sim \frac{4\pi^2\alpha' f}{\textrm{Vol}(B_3)^{1/2}}\nonumber
\end{align}
where $\Sigma_i\in H_{2}(S)$ and $\gamma_j\in H_{3}(B_3)$ denote the 2-cycles and 3-cycles that support the open and closed string fluxes, $n_i$ and $f_j$ are integer numbers, and the parameter $n$ ($f$) denote complex combinations of the various $n_i$ ($f_j$) and the complex structure moduli of $B_3$.  
In the same vein, the derivatives of $G(z,\bar z)$ scale as
\begin{equation}
G_y \sim \frac{2\, c_{y,G}}{\textrm{Vol}(B_3)^{1/6}}\ , \qquad G_{y\bar y} \sim \frac{4\, c_{y\bar y,G}}{\textrm{Vol}(B_3)^{1/3}}\ ,\label{deriva}
\end{equation}
where $c_{y,G}$ ($c_{y\bar y,G}$) is an adimensional complex (real)  constant. 

A comment on the scale of SUSY-breaking is in order here. Defining the scale of SUSY-breaking as $M_{\rm SS}\simeq m_{\tilde q}$, from the above expressions we have
\begin{equation}
M_{\rm SS}\sim 2\pi^2\alpha'|f|\left(\frac{g_s}{\textrm{Vol}(B_3)}\right)^{1/2}\sim \frac{|f|\, M_{\rm GUT}^2}{2\pi M_{\rm Pl}\sqrt{\alpha_{\rm GUT}}}
\end{equation}
Hence, for $|f|\simeq 1$ and standard unification values $M_{\rm GUT}\simeq 10^{16}$ GeV and $\alpha_{\rm GUT}^{-1}\simeq 24$ we would obtain $M_{\rm SS}\simeq 6.5\times 10^{12}$ GeV. However, it should be noted that the parameter $f$ in general receives contributions from a large number of 3-cycles (c.f.~eq.~(\ref{F})), so that large cancellations can take place that lead to $|f|\ll 1$ and lower the scale of SUSY-breaking. 
This is similar to what occurs for the small superpotential parameter $W_0$ in KKLT vacua \cite{Kachru:2003aw} and more generally for the cosmological constant in flux compactifications \cite{Bousso:2000xa, Sumitomo:2012wa, Sumitomo:2012vx, Sumitomo:2012cf}. In this work we therefore take $|f|$ (and thus $M_{\rm SS}$) to be a tunable parameter, perhaps selected on anthropic grounds.
Nevertheless, the reader may note that the dependence on $f$ of the flavor mixing parameters $\delta_{ij}$ and $\rho_{ij}$ cancels, and thus the limits that we obtain below for the sfermion masses do not actually depend on the tuning of $f$.

Making use of the scalings (\ref{F}) and (\ref{deriva}) we can estimate the dependence of the non-universal soft scalar masses on the local expansion parameter $\varrho$ defined in eq.~(\ref{decouple}). Indeed, from eqs.~(\ref{mii}) and (\ref{mij}) we observe that $(\delta m_{22}^2-\delta m_{11}^2)$ is of order $\mathcal{O}\left(\varrho^2\right)$ 
while the off-diagonal mass $m_{12}^2$ is of order $\mathcal{O}\left(\varrho\right)$ and is thus dominant. For the two lightest families we have 
\begin{equation}
\tilde\delta_{12}^{\rm d}=\frac{\tilde m_{12}^2}{\tilde m_{\tilde q}^2}\ =\ \frac{\sqrt{2}G_y} {\sqrt{q|M_0|}}\ \sim\ \frac{2\, c_{y,G}}{\sqrt{\pi |n|}}  \left(\frac {5} {3}\right)^{1/4} \varrho
\label{difmas}
\end{equation}
In particular, when $\varrho \ll 1$ the closed string flux varies very little over $S$ and the 
 flavor dependence is suppressed, as expected. 
In terms of physical scales
\begin{equation}
\tilde\delta_{12}^{\rm d}=\frac{\tilde m_{12}^2}{\tilde m_{\tilde q}^2}\ \sim\ 1.4\times \frac {c_{y,G}}{\sqrt{|n|}}\, \left(\frac {M_{\rm GUT}}{M_{\rm Pl}\,\alpha_{\rm GUT}}\right)^{1/3} \ .
\label{deltafinalG}
\end{equation}

The same estimate applies also to the off-diagonal flavor transitions for sleptons, parametrized by $\tilde \delta^{\rm l}_{12}$. Note however that the effective local magnetization parameter $n$ is expected to differ for sleptons and squarks, as they are differently charged under the hypercharge flux that breaks SU(5) down to the SM gauge group \cite{Donagi:2008ca, Beasley:2008kw, Donagi:2008kj}. This effect is also responsible for the breaking of the degeneracy between down-quark and lepton Yukawa couplings \cite{Aparicio:2011jx}. Making use of the numerical estimates of \cite{Aparicio:2011jx} for the local density of hypercharge flux one may check that the effect of the latter on the off-diagonal soft masses is small and, within the model-dependent uncertainties of the above estimation, we can take $\tilde \delta^{\rm d}_{12}\sim \tilde \delta^{\rm l}_{12}$ at the unification scale. Thus, the difference in the limits below for sleptons and squarks comes mainly from their different running under the RG.

Note from eq.~(\ref{deltafinalG}) that off-diagonal soft masses turn out to be parametrically suppressed by the ratio between the unification and the Planck scales. 
In a
standard unification scheme with $M_{\rm GUT}\simeq 10^{16}$ GeV this suppression is only of order $\sim 0.1$ and the actual value at $M_{\rm GUT}$ therefore depends substantially on the particular details of the 
magnetization parameter $n$ and the closed string flux variation parameter $c_{y,G}$, allowing for relatively large amounts of non-universal soft masses at the unification scale. On the other hand, as we have discussed in the beginning of this section, at low-energies there is an extra suppression factor
$(1+g(t)\xi^2)^{-1}$ that comes from the RG running between the unification scale and the SUSY-breaking scale $M_{\rm SS}$ and that dilutes the non-universal terms substantially in the case of squarks and more weakly in the case of sleptons.

\begin{figure}[t]
\begin{center}
\includegraphics[width=0.492\textwidth]{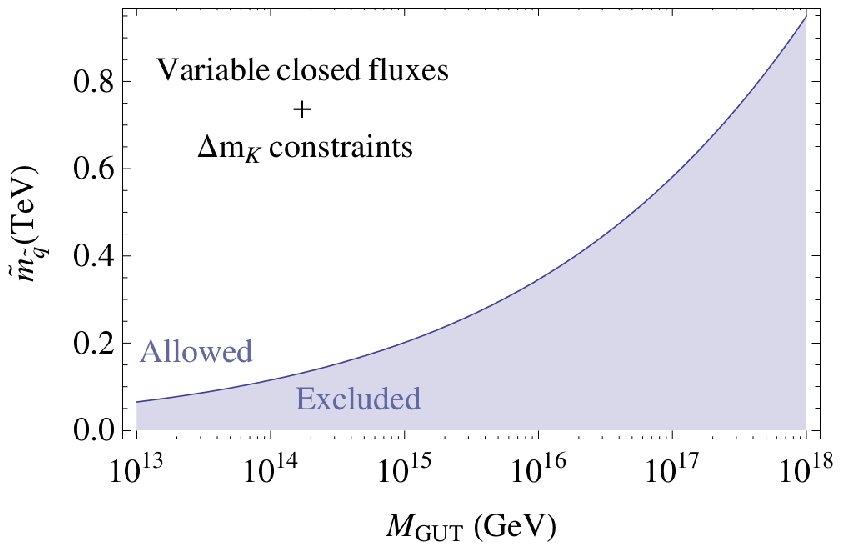} \
\includegraphics[width=0.492\textwidth]{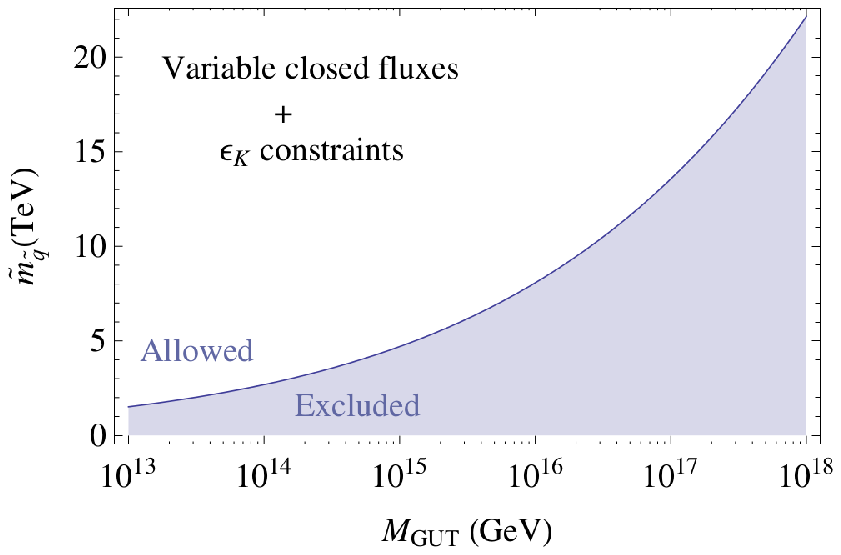}
\end{center}
\caption{\label{gcu}  Left: Constraints on squark masses $\tilde m_{\tilde{q}}$ vs the unification scale $M_{\rm GUT}$ coming from the kaon mass difference $\Delta m_K$ induced by non-constant densities of closed string fluxes along the GUT 4-cycle.
Right:  Analogous constraints coming from the CP violation parameter $\epsilon_K$.}
\label{closedlimits}
\end{figure}

Measurements of the kaon mass difference $\Delta m_K$ put constraints on the real part of the mixing $\tilde \delta_{12}^{\rm d}$ at the low-energy scale. The current experimental bounds require \cite{Giudice:2008uk} (see also \cite{Mescia:2012fg})
\begin{equation}
\left|\textrm{Re }{\tilde \delta}_{12}^{\rm d}\right| \ =\ 
\frac{\tilde m_{12}^2}{\tilde m_{\tilde{q}}^2}\ <\ 4.2\times 10^{-2}\ \frac{\tilde m_{\tilde{q}}}{350 \textrm{ GeV}}
\label{delta}
\end{equation}
where $\tilde m_{\tilde q}$ is the averaged squark mass at the scale $M_{\rm SS}$.
From eq.~(\ref{deltafinalG}) we then get a lower bound on the averaged squark mass
\begin{equation}
\tilde m_{\tilde{q}}\ \gtrsim\  \frac{\left|\textrm{Re } c_{y,G}\right|}{(1+g(t)\xi^2)\sqrt{|n|}}\times \left(\frac{M_{\rm GUT}}{M_{\rm Pl}}\right)^{1/3}\times 35 \textrm{ TeV}
\label{mq1}
\end{equation}
where we have taken $\alpha_{\rm GUT}^{-1}=24$. Note that for typical values $M_{\rm GUT}\simeq 10^{16}$ GeV, $n\simeq c_{y,G}\simeq 1$ and $\xi^2\simeq 2$, this expression leads to $\tilde m_{\tilde q}\gtrsim 350$ GeV,  
which is in the range already excluded by direct LHC  searches.  Thus $\Delta m_K$ does not provide strong bounds on the non-universalities induced by non-constant densities of closed string fluxes.
This is also shown in figure \ref{closedlimits} (left) where we have represented the low-energy bound on the averaged squark mass as a function of the unification scale. The bound becomes weaker as the unification scale is lowered, due to the decreasing of the flux variation over $S$.

The situation becomes much tighter if we consider the experimental constraints that come from the 
measured CP violation parameter $\epsilon_K$.  These yield \cite{Giudice:2008uk}
\begin{equation}
\left|\textrm{Im } {\tilde \delta}_{12}^{\rm d} \right|  \ <\ 1.8\times 10^{-3}\ \frac{\tilde m_{\tilde{q}}}{350 \textrm{ GeV}} \ .
\label{idelta}
\end{equation}
The local density of closed string flux $G(z,\bar z)$ is complex and so is the parameter $c_{y,G}$. We therefore expect the
real and imaginary parts of $\tilde m_{12}$ to be generically of the same order. In that case, the strong constraints coming from $\epsilon_K$ translate into
a more stringent limit for the averaged squark mass
\begin{equation}
\tilde m_{\tilde{q}}\ \gtrsim\ \frac{\left|\textrm{Im } c_{y,G}\right|}{(1+g(t)\xi^2)\sqrt{|n|}}\times \left(\frac{M_{\rm GUT}}{M_{\rm Pl}}\right)^{1/3}\times 8.1\times 10^2 \textrm{ TeV}
\label{mqcp1}
\end{equation}
A standard unification scale $M_{\rm GUT}\simeq 10^{16}$ GeV and typical values $n\simeq c_{y,G}\simeq 1$ and $\xi^2=2$ imply an averaged squark mass of at least $ \tilde m_{\tilde q} \gtrsim 8$ TeV, well above the current bounds coming from direct searches at the LHC.  Thus, while we have seen that the constraint on the real part of $\tilde m_{12}^2$ is not very relevant in this context, the constraint on the imaginary part turns out to be much stronger.  We have depicted the bound on the averaged squark mass versus the unification scale in figure \ref{closedlimits} (right).

The limits coming from the (yet unmeasured) $\mu \rightarrow e \gamma$ decay turn out to be quite strong. Using results from \cite{Arana-Catania:2013nha} we can obtain
an estimate based on recent experimental data \cite{Adam:2013mnn}
\begin{equation}
|\tilde \delta_{12}^{\rm l}|\ <\ 4\times 10^{-4}\ \left(\frac{\tilde m_{\tilde l}}{500\textrm{ GeV}}\right)^2 \ . \label{sleptonlim}
\end{equation}
From the leptonic analogue of eq.~(\ref{deltafinalG}) we then obtain
\begin{equation}
\tilde m_{\tilde{l}}\ \gtrsim\  \sqrt{\frac{\left|c_{y,G}\right|}{(1+g(t)\xi^2)|n|^{1/2}}}\times \left(\frac{M_{\rm GUT}}{M_{\rm Pl}}\right)^{1/6}\times 51 \textrm{ TeV} \ .
\end{equation}
If $M_{\rm GUT}\simeq 10^{16}$ GeV,  $\xi^2=2$ and $c_{y,G}\simeq n\simeq 1$  then $\tilde m_{\tilde{l}}\, \gtrsim\, 11$ TeV.
This is a remarkably strong bound which, of course, may be substantially weakened by playing with the model-dependent 
parameters, but it suggests that large masses for sleptons  are generically required.

\subsection{Flavor mixing  from non-constant open string fluxes}

We now study the non-universalities that arise from non-constant local densities of open string fluxes. For that aim, we consider the same $\bar{\mathbf{5}}$ matter curve of the previous sections, in this case with a constant density of closed string 3-form flux $G_0$ and a non-constant density of magnetization $M(z,\bar z)$ along the 4-cycle $S$. In order to evaluate (\ref{massfinal}) we proceed in the same way as we did for non-constant densities of closed string fluxes. Thus, we expand the density of magnetization around the point $x=y=0$ where the wavefunctions (\ref{wave1}) localize
\begin{equation}
M(z,\bar z) =M_0\left(1+M^*_yy+M_y\bar{y}+M_{y\bar y}|y|^2+\ldots\right)
\label{f2exp}
\end{equation}
We have not displayed the expansion on $x$ since it plays no role in the generation of non-universalities, as it will become clear below.
 
In order to obtain the soft scalar mass matrix we must evaluate (\ref{massfinal}) on this background, namely
\begin{multline}
m^2_{ij}= \frac{g_s}{4}\int_{\mathbb{C}^2}d^2xd^2y|G_0|^2\left[1-\left|\frac{M_0}{m}\right|\left(1+M_y^*y+M_y\bar{y}+
M_{y\bar y}|y|^2+\ldots\right)\right]\psi^+_i(\psi^+_j)^*
\label{intmag}
\end{multline}
Note that the local Gaussian wavefunctions introduced in section \ref{sec2} were actually derived for constant open string fluxes and are in principle not directly applicable to this case. However, it was shown in \cite{Font:2009gq} that the wavefunctions for non-constant open string fluxes have the same Gaussian
structure (\ref{psipm}) multiplied by a polynomial expansion on the local variables $x$, $\bar x$, $y$ and $\bar y$. Thus, to first order in the coordinate expansion the effect of this factor is to contribute  a further term, linear on $y$, on the integrand. This in practice amounts to a redefinition of the coefficient $M_y$ in (\ref{f2exp}). Therefore, in what follows $M_y$ represents an effective coefficient that includes not only the effect from the varying flux density $M(z,\bar z)$ but also the correction from
the modified wavefunction.

The integral (\ref{intmag}) with the wavefunctions (\ref{psipm}) is formally equivalent to (\ref{m2}), so that we can borrow the results of the previous subsection to obtain
\begin{align}
m^2_{\tilde q}&=\frac{g_s}{4}|G_0|^2\left(1-\left|\frac{M_0}{m}\right|\right)\ , & \delta m_i^2&=\frac{g_s}{4} |G_0|^2\frac{M_{y\bar y}}{q|m|}(4-i)\label{magmm}\\
m_{12}^2&=\frac{g_s}{4}|G_0|^2\frac{M_y}{|m|}\sqrt{\frac{2|M_0|}{q}}\ , & m_{23}^2& =\frac{g_s}{4}|G_0|^2\frac{M_y}{|m|}\sqrt{\frac{|M_0|}{q}}
\nonumber
\end{align}
where we have organized the soft scalar mass matrix accordingly to eq.~(\ref{matrix}). To estimate the size of the non-universalities we note in addition to eqs.~(\ref{F}) that the coefficients of the magnetization expansion scale with the volumes as 
\begin{equation}
M_y \sim \frac{2\, c_{y,F}}{\textrm{Vol}(S)^{1/4}}\ , \qquad M_{y\bar y} \sim \frac{4\, c_{y\bar y,F}}{\textrm{Vol}(S)^{1/2}}\ ,
\end{equation}
where $c_{y,F}$ and $c_{y\bar y,F}$ are adimensional complex constants. Moreover, the parameter $m$ in the background for $\Phi$, eq.~(\ref{phiback}), scales as (see e.g. \cite{Marchesano:2010bs})
\begin{equation}
m\simeq \frac{\eta}{2\pi\alpha'}
\label{laeta}
\end{equation}
where $\eta$ is a complex adimensional parameter related to the angle of the intersection between the GUT branes and the extra U(1) D7-brane. Plugging these scalings into eqs.~(\ref{magmm}) we obtain
\begin{gather}
\delta^{\rm d}_{12}\ =\ \frac{m_{12}^2}{m^2_{\tilde q}}\ \sim\ \frac{8\pi^{3/2}\, c_{y,F}\, |n|^{1/2}}{\eta}   \left(\frac {5} {3}\right)^{1/4}
\frac{\alpha'}{\textrm{Vol}(S)^{1/2}}\\ 
\rho^{\rm d}_{12}\ =\ \frac{\delta m_{2}^2-\delta m_1^2}{2m^2_{\tilde q}}\ \sim\ -\frac{4\pi\, c_{y\bar y,F}}{\eta}\left(\frac {5} {3}\right)^{1/2}\frac{\alpha'}{\textrm{Vol}(S)^{1/2}}\nonumber
\label{difmas3}
\end{gather}
Thus, unlike the case of varying closed string fluxes, for a non-constant density of open string flux along $S$, both $\delta_{12}^{\rm d}$ and $\rho_{12}^{\rm d}$ are parametrically of the same order. Making use of eq.~(\ref{d12til}) and expressing the result in terms of the physical scales we have that in the basis where the quarks mass matrix is diagonal, $\tilde \delta_{12}^{\rm d}$ reads
\begin{equation}
{\tilde \delta}_{12}^{\rm d}\ \sim\ \frac{1}{\eta}\left(\frac{M_{\rm GUT}}{M_{\rm st}}\right)^{2}\left(1.28\cdot c_{y,F}\sqrt{|n|} \cos 2\theta - 0.41\cdot  c_{y\bar y,F}\sin 2\theta\right)
\label{dmMx}
\end{equation} 
where $M_{\rm st}=\alpha'^{-1/2}$ is the string scale.

These results show  that there are potentially sizeable off-diagonal transitions, which are parametrically suppressed by $M_{\rm GUT}/M_{\rm st} = (2\alpha_{\rm GUT}/g_s)^{1/4}$. Barring fine-tunings, for generic $\theta$ 
 the second contribution in (\ref{dmMx}) is somewhat smaller than the first  for $n\geq 1$, so that we use the first term for our estimation of squark limits. From the constraints on $\Delta m_K$ discussed in the previous subsection we obtain that the averaged squark mass is bounded from below as
\begin{equation}
\tilde m_{\tilde q} \ \gtrsim \ \frac{1}{1+g(t)\xi^2}\sqrt{\frac{|n|}{g_s}}\frac{c_{y,F}}{\eta} \times 3.1 \ \textrm{TeV} \ .
\label{mqD}
\end{equation}
In particular, for $g_s\simeq n \simeq c_{y,F}\simeq \eta \simeq 1$ and $\xi^2=2$,  we get  $\tilde m_{\tilde q}\gtrsim 330$ GeV,  
and as in the closed string flux case, the constraints are weaker than the direct limits from  LHC. 

The bound  for the imaginary part of $\tilde \delta_{12}^{\rm d}$ coming from the CP violation parameter $\epsilon_K$ is stronger and gives
\begin{equation}
\tilde m_{\tilde q} \ \gtrsim \ \frac{1}{1+g(t)\xi^2}\sqrt{\frac{|n|}{g_s}}\frac{c_{y,F}}{\eta} \times 72 \ \textrm{TeV} \ .
\end{equation}
so that for $g_s\simeq n \simeq c_{y,F}\simeq \eta \simeq 1$ and $\xi^2=2$ we get a lower bound $\tilde m_{\tilde q}\gtrsim 7.6$ TeV for the squarks,
similar to the bound that arises for non-constant closed string flux densities.

\begin{figure}[!ht]
\includegraphics[width=0.5\textwidth]{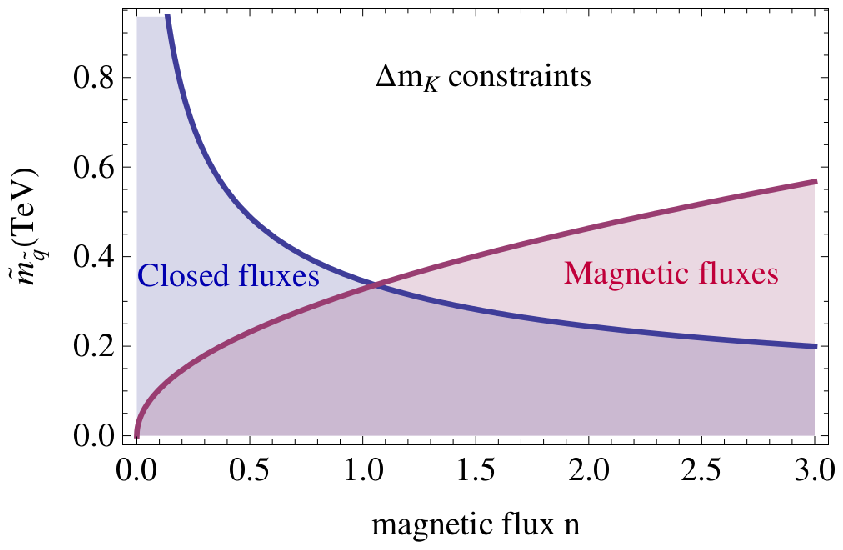}
\includegraphics[width=0.5\textwidth]{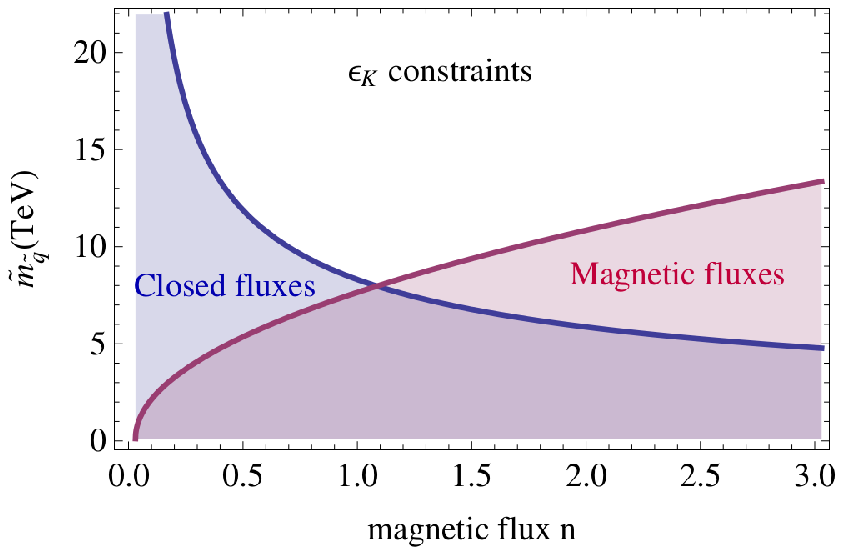}
\vspace{-15pt}
\renewcommand{\figurename}{\footnotesize{\textsc{Figure}}}
\caption{\footnotesize{Bounds on the averaged squark mass due to non-constant closed and open flux densities along the 4-cycle $S$  if $M_{\rm GUT}\simeq 10^{16}$ GeV. Left: bound coming from kaon mass mixing parameter 
$\Delta m_K$. Right: bound coming from the kaon CP violation parameter $\epsilon_K$.}}
\label{boundstodas}
\end{figure}

We show in figure \ref{boundstodas}  the lower averaged squark mass bound from  closed and open string fluxes as a function of the magnetization parameter $n$, coming 
from both $\Delta m_K$ and $\epsilon _K$. Combining closed and open string fluxes, squarks should be 
heavier than $\tilde m_{\tilde q}\gtrsim 8$ TeV to suppress sufficiently the contribution to CP violation in the kaon system.

It is also illuminating to look at the dependence of these bounds on the unification scale. Note in particular that the ratio $M_{\rm GUT}/M_{\rm st}$ in eq.~(\ref{dmMx}) can be fixed in terms of $g_s$ and $\alpha_{\rm GUT}$. Thus, for fixed $g_s$ and $\alpha_{\rm GUT}$ the non-universal corrections from the magnetization do not have a direct dependence on the value of the unification scale. Despite of this, an indirect dependence appears due to the renormalization of squark masses from $M_{\rm GUT}$ down to low-energies. The higher the value of $M_{\rm GUT}$, the  larger is the effect of the running and
the induced non-universalities are further diluted.
\begin{figure}[!ht]
\begin{center}
\includegraphics[width=0.6\textwidth]{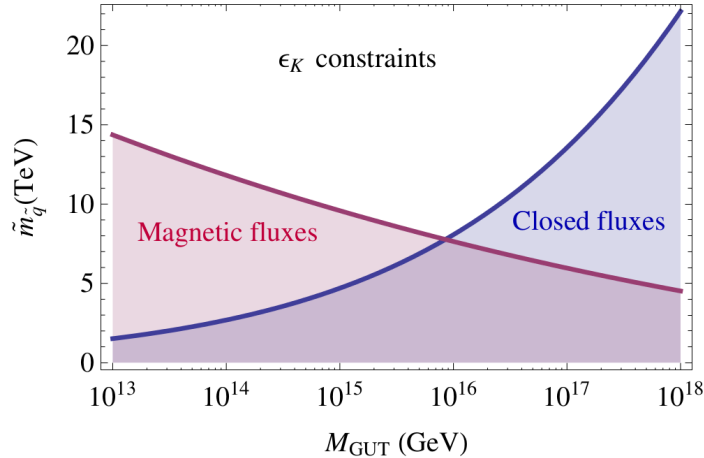}
\end{center}
\vspace{-15pt}
\renewcommand{\figurename}{\footnotesize{\textsc{Figure}}}
\caption{\footnotesize{Lower bound on the squark mass from CP violation as a function of the unification scale $M_{\rm GUT}$.
The contribution of both closed and open string fluxes is shown.}}
\label{Msdepen}
\end{figure}
This indirect dependence on $M_{\rm GUT}$ is of course also present for the non-universal terms
induced by closed string fluxes, but in that case the direct dependence on $(M_{\rm GUT}/M_{\rm Pl})^{1/3}$, c.f. eq.~(\ref{deltafinalG}), dominates over the indirect dependence. As a consequence the bounds on the averaged squark mass decrease in this case for lower $M_{\rm GUT}$.
This is displayed  in figure \ref{Msdepen} for the limits that come from the CP violation parameter $\epsilon_K$.
Interestingly, the weakest bounds are obtained for the standard values of the unification scale $M_{\rm GUT}\simeq 10^{16}$ GeV suggested by gauge coupling unification.

Let us finally turn to the limits that come from the yet unobserved branching ratio  BR$(\mu\rightarrow e \gamma)$.  From eq.~(\ref{sleptonlim}) and the leptonic analogue of (\ref{dmMx}) one obtains
\beq
\tilde m_{\tilde l} \ \gtrsim \ \frac{1}{\sqrt{1+g(t)\xi^2}}\left(\frac{|n|}{g_s}\right)^{1/4}\sqrt{\frac{\left|c_{y,F}\right|}{\eta}} \times 15.2 \ \textrm{TeV}
\eeq
If $M_{\rm GUT}\simeq 10^{16}$ GeV then one gets $\tilde m_{\tilde{l}}\, \gtrsim\, 10.6$ TeV, quite similar to the results
obtained from non-constant closed string fluxes.
\begin{figure}[!ht]
\begin{center}
{\includegraphics[width=0.49\textwidth]{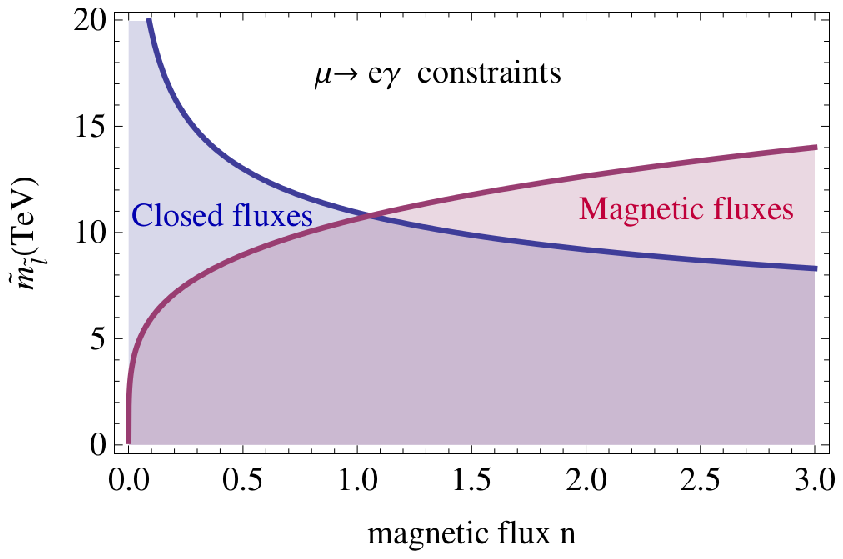}}\ \ 
{\includegraphics[width=0.49\textwidth]{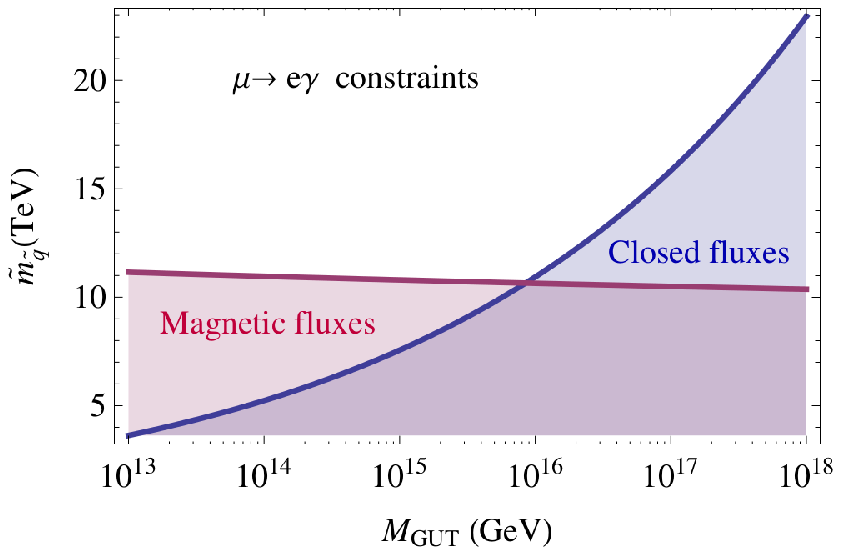}}
\end{center}
\vspace{-15pt}
\renewcommand{\figurename}{\footnotesize{\textsc{Figure}}}
\caption{\footnotesize{Bounds on the average slepton masses as a function open string flux $n$ and unification 
scale $M_{\rm GUT}$. Contributions from non-constant open and closed string fluxes are shown.\label{boundsleptons}}}
\end{figure}
\begin{table}[!ht]
\begin{center}
	\begin{tabular}{|c||c|c|c|}
\hline
	 & At $M_{\rm GUT}$ & At \ TeV\  scale & Experimental\  constraint \\
\hline
\hline
		{$\left|\text{Re }  \tilde\delta_{12}^{\rm d}\right|$} & \begin{tabular}{c} $0.39\ a$\\ $0.37\ b$\end{tabular} & \begin{tabular}{c}$4.1\times 10^{-2}\ a$\\ $3.9\times 10^{-2}\ b$\end{tabular} &  $4.2\times 10^{-2}\ \dfrac{\tilde m_{\tilde q}}{350\textrm{ GeV}}$ \\
\hline
	        {$\left|\text{Im } \tilde\delta_{12}^{\rm d}\right|$} & \begin{tabular}{c}$0.39\ a$\\ $0.37\ b$\end{tabular}   & \begin{tabular}{c} $4.1\times 10^{-2}\ a$\\ $3.9\times 10^{-2}\ b$\end{tabular} & {$1.8\times 10^{-3}\ \dfrac{\tilde m_{\tilde q}}{350 \textrm{ GeV}}$} \\
\hline
		{$\left|\tilde\delta_{12}^{\rm l}\right|$} & \begin{tabular}{c}$0.39\ a$\\ $0.37\ b$\end{tabular}  & \begin{tabular}{c} $1.9\times 10^{-1}\ a$\\ $1.8\times 10^{-1}\ b$\end{tabular}  &     $4\times 10^{-4}\ \left(\dfrac{\tilde m_{\tilde l}}{500\textrm{ GeV}}\right)^2 $   \\
\hline
	\end{tabular}
\end{center}
\caption{Prediction and experimental constraints for the mixing parameters $\delta_{12}$ coming from $\Delta m_K$, $\epsilon_K$ and BR($\mu\rightarrow e\gamma$) measurements/limits. The model-dependent parameters $a$ and $b$ are defined as $a=c_{y,G}/|n|^{1/2}$ and $b=c_{y,F}|n|^{1/2}/(g_s^{1/2}\eta)$ and are expected to be of order $\sim 1$.\label{tablafin}}	
\end{table}

In figure \ref{boundsleptons}
we show a summary of the bounds on the selectron and smuon masses for different values of $n$ and $M_{\rm GUT}$.
Note that in settings like this, where slepton and squark masses unify at $M_{\rm GUT}$, having sleptons with masses of order $\sim 10$ TeV would imply 
much heavier squarks, with masses as large as $\sim 25$ TeV,  quite above the bounds coming from the kaon system.

As a general summary of the numerical results obtained in this section for varying closed and open string fluxes, we present in table \ref{tablafin}
the expected values of   the real and imaginary parts of 
$\tilde\delta_{12}^{\rm d}$ and of $\tilde\delta_{12}^{\rm l}$, both at the $M_{\rm GUT}$ and the TeV scales,
as well as the corresponding experimental limits.

\section{Flavor violation, symmetries and the LHC reach}
\label{sec4}

Given the stringent results obtained  above, with squark and slepton masses  above the $\sim 10$ TeV range,
inaccessible to LHC, a natural question arises.  Under what conditions these bounds may be released and allow
for a SUSY spectrum within experimental reach at LHC?

The answer is obviously that those compactifications in which the fluxes vary very slowly on $S$ will 
get relaxed bounds.  Examples of such models are the toroidal  Type IIB orientfolds (and orbifolds there-off) of e.g.~\cite{Cascales:2003zp,Marchesano:2004xz,Marchesano:2004yq,Font:2004cx}
in which indeed constant fluxes are used. On the other hand one may reasonably argue that such models 
are not completely realistic and in any event rather un-generic.

It could be argued though  that 
we have to some extent assumed  the most pessimistic scenario in which the variation of  fluxes within the
manifold $S$ is linear in the local coordinates with coefficients $c_{y,G}$ and $c_{y,F}$ of order one. 
The bounds on squark masses are proportional to  these coefficients so that if for some reason they are suppressed
(say, $ |c_{y,G}|\simeq |c_{y,F}|\simeq 1/4$) squarks could be accessible to LHC. One possibility is that 
some symmetry (e.g.~$x,y\rightarrow -x,-y$) forbids the linear variation of the fluxes,
see e.g.~\cite{Kobayashi:2006wq, Abe:2009vi, Abe:2010ii, Nilles:2012cy, BerasaluceGonzalez:2012vb, BerasaluceGonzalez:2012zn, Berasaluce-Gonzalez:2013sna, Marchesano:2013ega, Honecker:2013hda} for some recent papers on discrete flavour symmetries in string compactifications. In this case the first terms contributing to the 
flux expansion would be quadratic.  We can repeat the analysis in this  case to find, for non-constant closed string flux
density a contribution to the mass difference (contributing through eq.~(\ref{d12til}) to $\tilde \delta _{12}^{\rm d}$)
\beq
\rho_{12}^{\rm d}\ =\ \frac{\delta m^2_{2}-\delta m_{1}^2}{2\tilde m_{\tilde q}^2}\ =\ \frac{|G_{y\bar y}|}{2q|M_0|}\ =\ \frac{|c_{y\bar y,G}|}{\pi |n|}\left(\frac{5}{3}\right)^{1/2} \varrho^2
\ =\ 0.5\ \left|\frac{c_{y\bar y,G}}{n}\right|\left(\frac{M_{\rm GUT}}{M_{\rm Pl}\, \alpha_{\rm GUT}}\right)^{2/3}
\label{rho12c}
\eeq
that leads to the lower bound 
\beq
\tilde m_{\tilde{q}}\ \gtrsim\  \frac{\left|c_{y\bar y,G}\right|}{(1+g(t)\xi^2)|n|}\times \left(\frac{M_{\rm GUT}}{M_{\rm Pl}}\right)^{2/3}\times 35.9 \textrm{ TeV} \ .
\eeq
For  $M_{\rm GUT}\simeq 10^{16}$ GeV and parameters of order one, one gets a very weak bound with  $\tilde m_{\tilde{q}}\, \gtrsim\, 34$ GeV.
Note that to quadratic order there is no imaginary contribution to $\delta_{12}^{\rm d}$ so there is no contribution to CP violation from this
correction.   Similar results are obtained from
the quadratic term coming from non-constant open string fluxes which yield
\beq
\rho_{12}^{\rm d}\ =\ \frac{\delta m^2_{2}-\delta m_{1}^2}{2\tilde m_{\tilde q}^2}\ =\ \frac{M_{y\bar y}}{2q|m|}\ =\ \frac{4\pi |c_{y\bar y,F}|}{\eta}\left(\frac{5}{3}\right)^{1/2} \frac{\alpha'}{\textrm{Vol}(S)^{1/2}}
 \ =\ 0.41\ \frac{|c_{y\bar y,F}|}{\eta}\left(\frac{M_{\rm GUT}}{M_{\rm st}}\right)^{2}
\label{rho12o}
\eeq
and gives rise to
\beq
\tilde m_{\tilde q} \ \gtrsim \ \frac{1}{1+g(t)\xi^2}\frac{1}{\sqrt{g_s}}\frac{|c_{y\bar y,F}|}{\eta} \times 0.99 \ \textrm{TeV} \ .
\eeq
For  $M_{\rm GUT}\simeq 10^{16}$ GeV and parameters of order one, one gets again a  very weak bound with  $\tilde m_{\tilde{q}}\, \gtrsim\, 105$ GeV, 
since there is no CP violating contribution to the kaon system in this case. In fact one can also check that similar limits to these may be obtained from the
quadratic contribution to $\textrm{Im}\, \tilde \delta_{13}^{\rm d}$. In this case the quadratic contribution is
complex and a contribution to CP violation in the $B^0_d$ system exists. We skip this analysis here for simplicity.

Regarding the limits that come from the $\mu\rightarrow e \gamma$ rate, it is possible to check that one gets 
from the leptonic analogues of (\ref{rho12c}) and (\ref{rho12o}) the lower bounds
\begin{align}
&\tilde m_{\tilde{l}}\ \gtrsim\  \sqrt{\frac{\left|c_{y\bar y,G}\right|}{(1+g(t)\xi^2)|n|}}\times \left(\frac{M_{\rm GUT}}{M_{\rm Pl}}\right)^{1/3}\times 52 \textrm{ TeV} & \quad &\textrm{(closed string flux)}\\
&\tilde m_{\tilde{l}}\ \gtrsim\ \frac{1}{\sqrt{1+g(t)\xi^2}}\frac{1}{g_s^{1/4}}\sqrt{\left|\frac{c_{y\bar y,F}}{\eta}\right|} \times 8.6 \ \textrm{TeV} & \quad &\textrm{(open string flux)} \ . \nonumber
\end{align}
With  $M_{\rm GUT}\simeq 10^{16}$ GeV and parameters of order one, the resulting limits on the averaged slepton mass are $\tilde m_{\tilde{l}}\, \gtrsim\, 3.4$ TeV from the first and 
$\tilde m_{\tilde{l}}\, \gtrsim\, 6$ TeV from the second.  Altogether we see that the limits on squark masses are totally relaxed 
if linear terms are absent 
whereas those from lepton number violation are only somewhat released. This is mostly due to the fact that the RG dilution is 
larger for squarks than for leptons.  Additional uncertainties in factors could allow for lighter slepton masses, within reach of LHC, 
but certainly the slepton limits are harder to relax.

\section{Discussion}

We have shown how non-constant fluxes in large classes of string compactifications with a MSSM structure give rise to substantial
flavor violating SUSY-breaking soft terms. We have concentrated on a setting with an underlying local SU(5) unification within
type IIB/F-theory, but the results may easily be extended to other compactifications.

In our  setting   the gauge fields correspond to 7-branes wrapping a 4-dimensional manifold $S$ inside a 6-dimensional 
compact manifold $B_3$. Matter fields are localised at complex matter curves within $S$ 
and open string fluxes $F_2$ provide for  the required 4d chirality.
One  can locally determine the internal wavefunctions of these matter fields  in terms of the 7-brane equations of motion. One finds that
the SM generations have a Gaussian profile, with different generations peaking at  different points over the
matter curve.  In the presence of IIB closed string fluxes $G_3$, SUSY-breaking soft terms are generated for the
fields living in the matter curve and in the gauge sector. When these fluxes are approximately constant over $S$, 
the soft terms obtained are flavor independent and correspond to the usual modulus-domination soft terms for
fields with {\it modular weight} 1/2 obtained using a supergravity approach
\cite{Ibanez:2004iv,Lust:2004dn,Font:2004cx,Aparicio:2008wh}. Our results provide the first microscopic derivation
of this soft-term structure, whose phenomenology at LHC has been analyzed in refs.\cite{Aparicio:2008wh, Aparicio:2012iw, Aparicio:2012vk, Aparicio:2012ju}.

In the more generic case of non-constant open and closed string fluxes, the flux felt by each generation is different,
since their wave functions are peaked at different points on the matter curve. This may give rise to important 
flavor violation in the sfermion  mass matrices, which require  squark  and slepton masses  of the first two generations 
to be in the $\sim 10$ TeV range.  Particularly strong are the limits coming from the decay $\mu \rightarrow e \gamma$.

In some models the non-constant local flux density near the GUT 7-branes could be due to the backreaction of localized SUSY-breaking sources present in the compactification (for non-constant closed string fluxes) or non-perturbative effects (for non-constant open string fluxes). In those cases, our computation can be understood as the supergravity computation of the non-universal threshold corrections to the soft scalar masses that result from integrating the corresponding heavy open string and/or non-perturbative modes out.

Our analysis may be extended in different directions.  We have considered here the simplest structure from
ISD $(0,3)$ closed string fluxes. A computation of soft terms  for fields on matter curves for more general classes of closed string fluxes will be
presented elsewhere \cite{inprogress}. Still no relevant changes are expected from these more general backgrounds with respect to the flavor
changing transitions that we have discussed here. In a different vein, 
one can perform similar computations in other compactification schemes.
For example, very similar results are expected in the case of type IIA models  with intersecting D6-branes or their  relatives, models based on M-theory 
compactifications on manifolds with $G_2$ holonomy. In particular, the three generations of SU(5) $\mathbf{5}$-plets will arise at different intersections of
a SU(5) stack with a U(1) stack. These intersections happen at different points in compact space and soft terms induced by
(non-constant) closed string fluxes will be different for the different generations. Furthermore the intersection angles for the three generations
(which are T-dual to the open string fluxes in IIB) will be generically different. This is the dual of having non-constant open string fluxes in the IIB side.
All in all, the same structure that we find in a  IIB/F-theory  is expected in this other large class of compactifications.
Finally, we have concentrated in our analysis on flavor violating transitions from the first and second quark/lepton generations, since they 
give the strongest constraints. It would be interesting to do a more through analysis including the third generation 
fermions

Given the numerical uncertainties we cannot exclude squarks being discovered  at the LHC. In particular we have shown 
how e.g. symmetries can substantially relax the obtained  mass limits, although this relaxation seems more difficult in the 
case of the first two generations of sleptons. On the other hand a heavy SUSY spectrum seems to be preferred by the
observed mass of the Higgs $m_H\simeq 126$ GeV. Our results seem to go in the same direction.

 \vspace{1.5cm}

\bigskip

\centerline{\bf \large Acknowledgments}

\bigskip

\noindent We thank   M.~Arana-Catania, M.~J.~Herrero, F.~Marchesano,  A.~Masiero and A.~Uranga for useful discussions. P.~G.~C\'amara thanks the CERN Theory Group and the Bethe Forum ``Supersymmetry: tools meet models'' for hospitality and partial support during the completion of this work. 
This work has been supported by the ERC Advanced Grant SPLE under contract ERC-2012-ADG-20120216-320421; 
by  the grants FPA 2009-09017, FPA 2009-07908, FPA 2010-20807-C02, Consolider-CPAN (CSD2007-00042) from the MICINN, HEPHACOS-S2009
/ESP1473 from the C.A. de Madrid, AGAUR 2009-SGR-168 from the Generalitat de Catalunya and the contract ``UNILHC" PITN-GA-2009-237920 of the European Commission. We also thank the 
spanish MINECO {\it Centro de excelencia Severo Ochoa Program} under grant SEV-2012-0249.

\end{document}